\documentclass[aps,prd,preprint]{revtex4}

\usepackage{enumerate}
\usepackage{amsmath,amssymb}
\usepackage{bm}
\usepackage{amscd}            % for extensible arrows (e.g., limits)
\usepackage{graphicx}
\usepackage{hhline,multirow}  % for nicer tables
\usepackage{dcolumn}          % Align table columns on decimal point

\newcommand{\mc}{\multicolumn}

\begin{document}

\preprint{JLAB-THY-09-09-1105}

\title{New parton distributions from large-$x$ and low-$Q^2$ data}

\author{A.~Accardi$^{a,b}$,
	M.~E.~Christy$^a$,
	C.~E.~Keppel$^{a,b}$,
	W.~Melnitchouk$^b$,
	P.~Monaghan$^a$,
	J.~G.~Morf\'\i n$^c$,	
	J.~F.~Owens$^d$ \\ \ }
\affiliation{
$^a$\mbox{Hampton University, Hampton, Virginia 23668}	\\
$^b$\mbox{Jefferson Lab, Newport News, Virginia 23606}	\\
$^c$\mbox{Fermilab, Batavia, Illinois 60510}		\\
$^d$\mbox{Florida State University, Tallahassee, Florida 32306-4350}\\
\\}

% \date{\today}

\begin{abstract}

We report results of a new global next-to-leading order fit of parton
distribution functions in which cuts on $W {\rm \ and \ } Q$ are
relaxed, thereby including more data at high values of $x$.  Effects
of target mass corrections (TMCs), higher twist contributions, and
nuclear corrections for deuterium data are significant in the large-$x$
region.  The leading twist parton distributions are found to be stable
to TMC model variations as long as higher twist contributions are also
included.  The behavior of the $d$ quark as $x \to 1$ is particularly
sensitive to the deuterium corrections, and using realistic nuclear
smearing models the $d$-quark distribution at large $x$ is found to be
softer than in previous fits performed with more restrictive cuts.

\end{abstract}

\maketitle

%%%%%%%%%%%%%%%%%%%%%%%%%%%%%%%%%%%%%%%%%%%%%%%%%%%%%%%%%%%%%%%%%%%%%%%%%
\section{Introduction}
\label{sec:intro}

The distribution of up and down quarks in the proton is one of the most
fundamental characterizations of the structure of the ground state of QCD.
Several decades of accumulated data from a variety of hard scattering
processes, together with sophisticated, next-to-leading order (NLO) QCD
analyses, has produced a detailed mapping of the proton's parton
distribution functions (PDFs) over a large range of kinematics
\cite{Nadolsky,MSTW,ABKM,Ball:2009mk}.

A dominant role in this endeavor has been played by lepton-nucleon deep
inelastic scattering (DIS) data, which provide direct information on the
behavior of the quark PDFs, as well as on the gluon distribution via the
observed logarithmic scaling violations and the imposition of the
momentum sum rule on the PDFs.
However, the kinematics of many deep inelastic experiments limits the
coverage in Bjorken $x$, so that our knowledge of PDFs does not extend
uniformly over the entire $x$-range, especially so at large $x$.
The invariant mass squared of the produced hadronic system is given by 
\begin{equation}
W^2 = M^2 + Q^2 \left( \frac{1}{x}-1 \right)\, ,
\end{equation}
where $M$ denotes the mass of the target nucleon and $Q^2$ the photon
virtuality.  As $x \to 1$ at fixed $Q^2$, \ $W$ approaches values in
the nucleon resonance region.
This region may be treated using the concept of quark-hadron duality
\cite{MEK}, but this is beyond the scope of the present analysis.
In order to simultaneously stay in the region of large $W$ and access 
high values of $x$, one must therefore go to increased values of $Q^2$;
in practice, though, this is limited by the available beam energy.
Finally, the cross section falls rapidly as $x {\rm \ and \ }Q^2$
increase, so the data samples can become statistics--limited in this
region.

Knowledge of PDFs at large $x$, defined here to be $x \gtrsim 0.5$,
is important for a number of reasons.
Apart from its intrinsic value in providing a laboratory for studying
the flavor and spin dynamics of quarks, the large-$x$ region is unique
in allowing perturbative QCD predictions to be made for the $x$
dependence of PDFs in the limit $x \to 1$ \cite{pQCD}.
Furthermore, the $d/u$ quark ratio at large $x$ is very sensitive
to different mechanisms of spin-flavor symmetry breaking in the
nucleon \cite{MT_NP}.
Reliable knowledge of PDFs at large $x$ may also be important for
searches of new physics signals in collider experiments, where
uncertainties in PDFs at large $x$ and low $Q^2$ percolate through
$Q^2$ evolution to affect cross sections at smaller $x$ and larger
$Q^2$ \cite{Kuhlmann}.
This is especially true if the search involves a region where the
rapidity is large, where one is sensitive to products of PDFs evaluated
with one value of $x$ being small and the other large.
An example is provided by the planned PHENIX measurements of polarized
gluon distributions, $\Delta g$, at small $x$ by detecting $W$-bosons
at large rapidity: since $\Delta g$ is convoluted with large-$x$
unpolarized quark distributions the precision of the measurements will
depend on the precision to which the quark PDFs are known at large $x$.

Nuclear PDFs at large $x$ are also important in the analysis of
neutrino oscillation experiments such as T2K \cite{Itow:2001ee},
NO$\nu$A \cite{Ayres:2004js}, and DUSEL \cite{Raby:2008pd}.
A large part of the theoretical uncertainty is due to lack of precise
knowledge of the neutrino--nucleus interaction in the kinematics
between the DIS and resonance regions, as well as in the implementation
of the non-isoscalarity correction.
Better control of nuclear corrections at large $x$ and a precise
knowledge of the $d/u$ ratio will have a direct and measurable impact
on the interpretation of such experiments.
This analysis is also timely given the increasing base of large-$x$
measurements available, including, among others, DIS data from 
Jefferson Lab, and Drell-Yan and $W$-asymmetry data from Fermilab.

In order to improve our knowledge of PDFs in the large-$x$ region,
and to thereby reduce their errors, in this work we expand the coverage
in $x$ provided by the deep inelastic data by relaxing the $W>3.5$~GeV
and $Q>2$~GeV cuts traditionally used to limit the theoretical analysis
to leading twist.
This requires the development of the methodology for treating $1/Q^2$
suppressed terms such as target mass corrections and higher twist
contributions.
Furthermore, as one probes deeper into the large-$x$ region, nuclear
corrections must be included when dealing with nuclear targets, as
these become increasingly important as $x \to 1$.

It is {\em a priori} not obvious that the uncertainties on the PDFs will
be reduced by relaxing the cuts to include more data, since the various
new theoretical contributions will have uncertainties of their own which
could increase the resulting uncertainties on the extracted PDFs.
It is the purpose of this paper to explore precisely this question. 
While some previous attempts have been made in the literature to include
lower-$Q$ and lower-$W$ data in global fits and to explore the related
uncertainties \cite{Alekhin01,Alekhin03,Alekhin06}, in this work we
specifically focus on the large-$x$ region.

The outline of the paper is as follows.  Section~II summarizes the
theoretical issues which must be addressed in the region of large $x$,
while Sec.~III addresses the choice of data sets and the procedures
utilized in the fitting.  In Sec.~IV the results of the fits are
presented and discussed, and Sec.~V describes the resulting $d/u$ ratio. 
Our conclusions are presented in Sec.~VI.
In keeping with the convention adopted in previous CTEQ fits
\cite{Nadolsky,CTEQ6.1}, we shall refer to the new fit from this
analysis as ``CTEQ6X.''

%%%%%%%%%%%%%%%%%%%%%%%%%%%%%%%%%%%%%%%%%%%%%%%%%%%%%%%%%%%%%%%%%%%%%%%%%
\section{Theoretical issues at large $x$}
\label{sec:theory}

From a theoretical standpoint the large-$x$ region requires the
inclusion not only of leading twist PDFs, but also of contributions
suppressed by at least one power of $Q^2$, which include target mass
corrections and higher twist corrections.
The former are kinematic in origin and involve terms suppressed by
powers of $M^2/Q^2$; the latter are dynamical in origin and are
suppressed as powers of $\Lambda^2/Q^2$ at large $Q^2$, with $\Lambda$
measuring the scale of nonperturbative parton-parton correlations.
Furthermore, if a nuclear target is used, one must include nuclear
corrections which account for the difference between the PDFs in
free nucleons and those in nucleons bound in a nucleus.
These corrections are not suppressed at high $Q^2$, and are most
significant at large $x$.
Historically, to limit the effects of the target mass and higher twist
corrections it has been customary to restrict the data included in the
fits to large values of $Q$ and $W$.
In this section we discuss the implications of these cuts and the
theoretical tools needed to relax them.

% .......................................................................
\subsection{Kinematic cuts}

To avoid theoretical complications in PDF analyses at large $x$ it has
been common to limit the deep inelastic data used to those points which
satisfy $Q \geq 2~{\rm GeV \ and \ } W \geq 3.5 {\rm \ GeV}$. 
These combined cuts have the effect of severely limiting the number of
DIS data points in the large-$x$ region that may be used to constrain
PDFs in a global fit. 
However, as discussed in Refs.~\cite{Pumplin:2009sc,Martin:2003sk},
they do not completely eliminate the need for ${\cal O}(1/Q^2)$ power
corrections, whose neglect is a source of tension between several
DIS data sets.

Other hard scattering processes, such as the production of lepton pairs,
vector bosons, high-$p_T$ jets, and direct photons, involve a large
momentum scale analogous to $Q^2$ in deep inelastic scattering. 
The values are typically sufficiently large that one avoids target
mass corrections and higher twist contributions.
However, the relevant range in $x$ is typically set by
$x_T=2 p_T/\sqrt{s}$ for high-$p_T$ scattering or by
$M_{l^+l^-}/\sqrt{s}$ for the production of a lepton pair of mass
$M_{l^+l^-}$, and data for such processes rarely constrain the
large-$x$ behavior of PDFs (although some exceptions to this will be
discussed below).

Given all of the restrictions described above, one might well wonder
how it is that any constraints at all are provided on the possible
behavior of PDFs at large $x$.
The answer is that there are indirect constraints which are provided
by sum rules and the $Q^2$ evolution of the PDFs.
The momentum and number sum rules provide some constraints since
they involve integrations over the entire range of $x$.
However, since the PDFs fall off as powers of $(1-x)$, the large-$x$
region contributes only small amounts to these sum rules.
In addition, the DGLAP $Q^2$ evolution equations, used to calculate
the PDFs at values of $Q^2$ above that where the initial PDFs are
parametrized, have the property that the PDFs at high values of $x$ and
low $Q^2$ feed the behavior at lower values of $x$ and higher $Q^2$.
Hence, there are indirect constraints placed on the large-$x$ PDF
behavior by data in regions of lower $x$ and higher $Q^2$.

It is important to understand that when one uses a supplied
parametrization of PDFs in the large-$x$ region the results
represent extrapolations of the PDFs into regions where they
have not been directly constrained by data.
In order to reduce the size of the unconstrained region and check
the validity of the large-$x$ extrapolations obtained in standard
global fits, one needs to account for the neglected theoretical
corrections, which we discuss below.

% .......................................................................
\subsection{Nuclear corrections}
\label{ssec:nuclear}

To probe the $x$ dependence of the $d$ quark to the same level of
accuracy attained for the $u$ quark requires lepton scattering data
from a neutron target.
Since free neutron targets do not exist, the next best choice is to
utilize a deuterium target.
Because the deuteron is a very weakly bound nucleus, many analyses
which include proton and deuteron data make the assumption that it
can be treated as a sum of a free proton and neutron.
On the other hand, it has long been known from experiments on a range
of nuclei that a nontrivial $x$ dependence exists for ratios of
nuclear to deuteron $F_2$ structure functions.
These effects include nuclear shadowing at small values of $x$,
anti-shadowing at intermediate $x$ values, $x \sim 0.1$, a reduction
in the structure function ratio below unity for
$0.3 \lesssim x \lesssim 0.7$, known as the European Muon
Collaboration (EMC) effect, and a rapid rise as $x \to 1$ due to
Fermi motion.
Moreover, deviations from unity of the ratio of deuteron to proton
structure functions corrected for non-isoscalarity, as well as many
theoretical studies, strongly suggest the presence of nuclear effects
also in the deuteron.

In the absence of experiments with free neutron targets, the size of the
nuclear effects in the deuteron has yet to be determined empirically,
although expectations are that it should be somewhat smaller than for
heavy nuclei on the basis of its much lower binding energy.
Nevertheless, in the region $x \gtrsim 0.5$ the effects of nuclear
Fermi motion lead to a rapidly increasing ratio of deuteron to nucleon
structure function, which diverges as $x \to 1$.
Any high-precision analysis of the large-$x$ region must therefore
account for the nuclear effects if data on deuterium (or other nuclei)
are used in the fit.

The conventional approach to describing nuclear structure functions
in the intermediate- and large-$x$ regions is the nuclear impulse
approximation, in which the virtual photon scatters incoherently
from the individual bound nucleons in the nucleus.
There have been many formulations of nuclear structure functions
in the literature within this framework (see {\it e.g.}
Refs.~\cite{West,Jaffe,Bickerstaff,GST}).
In this analysis we adopt the weak binding approximation (WBA) of
Refs.~\cite{WBA,WBApol,KP,KMK}, where a covariant framework is used
to relate the nucleon and deuteron scattering amplitudes, which are
systematically expanded in powers of $\bm{p}/M$, with $\bm{p}$ the
bound nucleon momentum in the deuteron.
The deuteron structure function can then be written as a convolution
of the bound nucleon structure function and the distribution of
nucleons in the nucleus.
Because the struck nucleon is off its mass shell with virtuality
$p^2 = p_0^2 - \bm{p}^2 < M^2$, its structure function can in
principle depend on $p^2$, in addition to $x$ and $Q^2$.

Since the deuteron is weakly bound, with binding energy
$\varepsilon_D = -2.2$~MeV and average nucleon momenta of the order
$|\bm{p}| \sim 130$~MeV, the typical nucleon virtuality will be
$\sim 4\%$ smaller than the free nucleon mass, $M$.
One can therefore approximate the bound nucleon structure function
by its on-shell value, in which case the deuteron structure function
can be written, to order $\bm{p}^2/M^2$, as \cite{KP,KMK,AQV}
\begin{eqnarray}
F_2^d(x,Q^2)
&\approx& \sum_{N=p,n} \int_x^{y_{\rm max}} dy\
    f_{N/d}(y,\gamma)\ F_2^N\left(\frac{x}{y},Q^2\right)\, .
\label{eq:F2d}
\end{eqnarray}
Here $F_2^N$ is the nucleon (proton $p$ or neutron $n$) structure
function, and $f_{N/d}$ gives the light-cone momentum distribution
of nucleons in the deuteron.
The scaling variable $y = (M_d/M) (p \cdot q/ p_d \cdot q)$ is the
deuteron's momentum fraction carried by the struck nucleon,
where $q$ is the virtual photon momentum, and $p_d$ and $M_d$
are the deuteron four-momentum and mass.
The maximum value of $y$, to order ${\bm p}^2/M^2$, is given by
$y_{\rm max} = 1 + \gamma^2/2 + \varepsilon_d/M$, so that
in the Bjorken limit $y_{\rm max} \to 1.5 + \varepsilon_d/M$.
Relativistically the upper limit on $y$ is given by $M_d/M$,
which suggests that the nonrelativistic approximation breaks
down at large $y$.  A careful treatment of relativistic
corrections will be important as $x \gtrsim 1$; however, for the
$x$ range covered in this analysis a nonrelativistic treatment is
sufficient.
In the Bjorken limit the nucleon distribution function $f_{N/d}$
(also called the ``smearing function'') is a function of $y$ only
and is limited to $y \leq M_d/M$.
At finite $Q^2$, however, it depends in addition on
$\gamma = \sqrt{1 + 4 x^2 M^2/Q^2}$, which in the nucleus rest frame
coincides with the virtual photon ``velocity'' $|\bm{q}|/q_0$, with
significant consequences when fitting large-$x$ deuterium data
\cite{KMK}.  Furthermore, at finite $Q^2$ the limits of integration
in $y$ in general become $x$ and $Q^2$ dependent, although their
effects are negligible for $x \lesssim 0.85$ \cite{KP,AQV}.

The function $f_{N/d}$ is computed from the deuteron wave function,
and accounts for the effects of Fermi motion and binding energy 
of the nucleons in the nucleus, as well as kinematic $1/Q^2$
corrections \cite{KP,KMK,AQV}.  In our numerical calculations we use
several different nonrelativistic deuteron wave functions, based on
the Paris \cite{Paris}, AV18 \cite{AV18} and CD-Bonn \cite{CD-Bonn}
nucleon--nucleon potentials, but find that the differences are small
for $x \lesssim 0.85$.

Including explicit off-shell dependence in the bound nucleon
structure function leads to a two-dimensional convolution in terms
of $y$ and the nucleon virtuality $p^2$.
For small values of $|p^2 - M^2|/M^2 \ll 1$ the off-shell effects
can be incorporated into a generalization of Eq.~(\ref{eq:F2d}) in
which the nucleon distribution function depends explicitly on $x$,
in addition to $y$ and $\gamma$.
However, at larger $x$ higher-order corrections in $\bm{p}/M$ become
increasingly important, and the factorization of the nucleon and
nuclear functions inherent in the convolution approximation is
expected to break down \cite{MST2,MST1}.
In this case relativistic effects can be incorporated through an
additive correction, $F_2^d \to F_2^d + \delta^{\rm (off)} F_2^d$,
which arises from explicit $p^2$ dependence in the quark--nucleon
correlation functions and relativistic $P$-state components of the
deuteron wave function.  The latter have been computed, for instance,
within the relativistic spectator theory for the deuteron \cite{Gross}.

The off-shell dependence of the bound nucleon structure functions
computed in Ref.~\cite{MST2} has been estimated within a simple
quark--spectator model \cite{MST1} with the parameters fitted to
proton and deuteron $F_2$ data, and leads to a reduction of about
2\% of $F_2^d$ compared to the on-shell approximation.  
A simple parametrization of the relativistic and off-shell effects
in the model of Ref.~\cite{MST2}, relative to the total $F_2^d$,
is given by
\begin{eqnarray}
\frac{\delta^{\rm (off)} F_2^d}{ F_2^d}
&=& a_0\, (1 + a_1 x^{a_2}) (1 - [a_3 - x^{a_4}]^{a_5})\, ,
\label{eq:delta_off}
\end{eqnarray}
with $a_i = \{ -0.014, 3, 20, 1.067, 1.5, 18 \}$
at a scale of $Q^2 \approx 5$~GeV$^2$.
Note that this correction is applicable only for $x \gtrsim 0.2$,
and for the values of $x$ relevant to the current analysis one finds
$\delta^{\rm (off)} F_2^d / F_2^d \lesssim 1.5\%$.
Furthermore, because the off-shell nucleon--deuteron amplitude was
evaluated in Ref.~\cite{MST2} using wave functions derived from a
pseudoscalar $\pi N$ interaction \cite{Gross}, which is known to produce
large $P$-state contributions, the correction (\ref{eq:delta_off}) is
likely to provide an upper limit on the size of the relativistic,
convolution-breaking effects.

Finally, we note that while a few global PDF analyses
\cite{Alekhin01,Alekhin03,Alekhin06,Blumlein:2006be} have incorporated
nuclear effects in the deuteron using smearing functions similar to
those employed here, some studies \cite{Yang} have corrected deuteron
data using a parametrization \cite{Gomez} of the nuclear density model
\cite{density}.
This model is motivated by the observation that for heavy nuclei the
ratio of nuclear to deuterium structure functions scales with the
nuclear matter density \cite{GST,density}.
Assuming the density scaling extends all the way down to the deuteron,
one can obtain the ratio of deuteron to nucleon structure functions
from the ratio of iron to deuterium structure functions,
$F_2^d/F_2^N \approx 1 + 0.25\ (F_2^{\rm Fe}/F_2^d - 1)$, using the
empirical nuclear density for $^{56}$Fe and an {\it ansatz} for the
charge density of deuterium \cite{density,comment}.
Consequently the $F_2^d/F_2^N$ ratio displays a relatively large
depletion at $x \sim 0.6$ and a rise above unity which does not
set in until $x \sim 0.8$, significantly higher than typically
found in smearing models.
Unfortunately for light nuclei, and especially deuterium, it is
difficult to define physically meaningful nuclear densities
\cite{comment}, so that the application of the density model to
deuteron data inevitably suffers from ambiguities.
Furthermore, even for heavy nuclei the authors of Ref.~\cite{density}
caution that the density model should only be considered qualitative
beyond $x \approx 0.6$--0.7.

% .......................................................................
\subsection{Target mass corrections}
\label{ssec:tmc}

In the context of the operator product expansion (OPE) in QCD, deep
inelastic scattering is formulated through moments of structure
functions, which are related to forward matrix elements of local
operators.
The $x$ dependence of structure functions is formally reconstructed
from the moments via an inverse Mellin transform.
At large $Q^2$ the process is dominated by matrix elements of operators
of twist two, such as the quark bilinear $\bar\psi \gamma^\mu \psi$.
Operators which include insertions of covariant derivatives,
$\bar\psi \gamma^\mu D^{\mu_1} \cdots D^{\mu_n} \psi$,
do not alter the twist, but have matrix elements that enter as
$M^2/Q^2$ corrections to the ${\cal O}(1)$ terms \cite{Nachtmann,GP}.
These ``target mass corrections'' (TMCs), which are of purely kinematic
origin, need to be removed from the empirical structure function data
before information on the twist-two PDFs can be extracted.

One of the limitations of the OPE formulation of TMCs is the so-called
``threshold problem'', in which the target mass corrected structure
functions remain nonzero at $x \geq 1$.
This arises from the difficulty in consistently incorporating the
elastic threshold in moments of structure functions at finite $Q^2$,
resulting in nonuniformity of the $Q^2 \to \infty$ and $n \to \infty$
limits, where $n$ is the rank of the moment.
A number of attempts to address the unphysical $x \to 1$ behavior have
been made \cite{KP,Tung,Steffens}, although none of the proposed
approaches is free of additional assumptions or complications
\cite{Schienbein}.

Alternatively, using the collinear factorization (CF) formalism
\cite{EFP,Collins} one can avoid the threshold ambiguities from
the outset by formulating TMCs directly in momentum space.
Recently this formalism was applied \cite{AQ} to deep inelastic
structure functions at large $x$ at next-to-leading order, carefully
taking into account the elastic threshold to render the structure
functions zero above $x=1$.
However, in the handbag approximation, without introducing a suitable
jet function to account for the invariant mass of the final hadronic
state, leading order structure functions can still be nonzero at $x=1$
\cite{AQ}.
A simplified version of the CF formalism, which involves replacing
the Bjorken variable $x$ with the Nachtmann variable
$\xi=2x/\left(1+\sqrt{1+4x^2M^2/Q^2}\right)$, was proposed in  
Refs.~\cite{Aivazis,KretzerCF}.
At leading order this ``$\xi$-scaling'' prescription coincides
with the CF results.
Beyond leading order, however, this overestimates the CF results
by about 20--30\% \cite{AQ,AM}.

In this analysis we consider both the traditional OPE approach and
the collinear factorization formulations of TMCs for DIS reactions.
For deuterium data we first apply the TMCs to the nucleon structure
function, which is then convoluted through Eq.~(\ref{eq:F2d}) to obtain
the deuteron $F_2^d$.
Although there is some residual dependence on the TMC prescription
adopted, we will see in Sec.~\ref{sec:results} below that the resulting
PDFs are essentially independent of the choice once allowance is made
for dynamical higher twist and other power corrections.
For non-DIS data, such as Drell-Yan, $W$-asymmetry and jet production
in hadronic collisions, the target mass and
nuclear corrections should play a negligible role because of the
limited reach at large $x$ and the typically large momentum scales
involved, and are therefore neglected in our analysis.

% .......................................................................
\subsection{Dynamical power corrections}
\label{ssec:ht}

In the OPE framework, dynamical higher twist contributions to DIS
structure functions are associated with matrix elements of operators
involving multi-quark or quark and gluon fields, which give rise to
$1/Q^2$ or higher order corrections to structure functions.
As with the kinematic target mass corrections, these must be taken into
account in analyses of data at low $Q^2$ and especially at large $x$.
Because dynamical higher twists involve nonperturbative multiparton
interactions, it is notoriously difficult to quantify their magnitude
and shape from first principles.

The usual approach in analyses whose main aim is the extraction
of leading twist PDFs is either to parametrize the higher twist
contributions by a phenomenological form and fit the parameters
to the experimental data \cite{AKP07,Pumplin:2002vw}, or to extract
the $Q^2$ dependence by fitting it in individual bins in $x$
\cite{Alekhin03,Martin:2003sk,Virchaux,MRST_HT98,Blumlein:2008kz}.
Such an approach effectively includes contributions from multiparton
correlations (the true higher twist contributions) along with other
power corrections that are not yet part of the theoretical treatment
of DIS at low $Q^2$.
These include ${\cal O}(1/Q^2)$ contributions such as jet mass
corrections \cite{AQ} and soft gluon resummation \cite{Schaefer:2001uh},
as well as contributions which are of higher order in $\alpha_s$ but
whose logarithmic $Q^2$ behavior mimics terms $\propto 1/Q^2$ at low
virtuality \cite{Blumlein:2008kz,Alekhin:1999kt}.
For these reasons it is more appropriate to speak about {\it residual
power corrections} rather than higher twist corrections.
Nevertheless, we shall refer to all of these $1/Q^2$ suppressed
effects as ``higher twist'' corrections in order to conform with the
terminology frequently used in the literature, but keeping in mind
their possibly disparate origins.

We employ the commonly-used phenomenological form for the total
structure function,
\begin{eqnarray}
F_2(x,Q^2)
&=& F_2^{\rm LT}(x,Q^2)
    \left( 1 + \frac{C(x)}{Q^2} \right)\, ,
\label{eq:HT}
\end{eqnarray}
where $F_2^{\rm LT}$ is the leading twist component including TMCs,
and the coefficient of the $1/Q^2$ term is parametrized (in units
of GeV$^2$) as
\begin{align}
  C(x) = c_1\, x^{c_2} (1 + c_3 x)\, .
\label{eq:HTparams}
\end{align}
The parameter $c_1$ reflects the overall scale of the higher twist
corrections, while $c_2$ controls the well-known rise of the
coefficient $C(x)$ at large $x$; the parameter $c_3$ allows for
the possibility of negative higher twists at smaller $x$. 
For deuterium targets we use the same higher twist term $C(x)$
for the bound proton and neutron $F_2$ structure functions. 
In so doing we neglect higher twist contributions coming from parton
rescattering on the spectator nucleon \cite{Qiu:1988dn} due to the
small atomic number, which would slightly increase the deuteron's
dynamical higher twist contribution compared to that of the proton.
These assumptions may need to be relaxed in future analyses,
see {\it e.g.} Refs.~\cite{Blumlein:2008kz,AKL04}.

We stress the importance of explicitly including {\em both} the TMCs
{\it and} the higher twist corrections, which have very different
physical origin and can have very different $x$ dependence.
Since the higher twists are fitted phenomenologically, one could in
principle simulate TMC effects with an empirical power corrections
function with a sophisticated enough parametrization. 
We have considered several alternative forms for $C(x)$, but were
not able to find an efficient parametrization which embodies both
the higher twist and TMC effects and allows a good fit.
On the contrary, inclusion of TMCs in the LT structure function in
Eq.~\eqref{eq:HT} allows us to adopt a very economical 3-parameter
functional form for $C(x)$.

%%%%%%%%%%%%%%%%%%%%%%%%%%%%%%%%%%%%%%%%%%%%%%%%%%%%%%%%%%%%%%%%%%%%%%%%%
\section{Data sets and fitting}
\label{sec:data}

We perform NLO global PDF fits to proton and deuteron data from
inclusive deep inelastic fixed-target scattering experiments at
Jefferson Lab (JLab), SLAC, and the BCDMS and New Muon Collaborations
(NMC) at CERN;
$ep$ collider data from H1 and ZEUS;
E605 and E866 Drell-Yan data from $pp$ and $pd$ collisions;
$W$-lepton asymmetry data from CDF and D0;
$W$~asymmetry data from CDF;
and jet and $\gamma$+jet cross sections from D0.
The full list of data sets, together with a comparison to the data sets
included in the most recent CTEQ6.1 fits \cite{CTEQ6.1}, is provided in
Table~\ref{tab:datasets}.

\begin{table}[tb]
\centering
\caption{Data sets and number of data points (total and deuterium)
	used in the global fits discussed in the text. The data sets
	with a check mark ($\surd$) were used in the CTEQ6.1 fits
	\cite{CTEQ6.1}.\\ } 
\begin{tabular}[c]{lll|cc|cc|c}  \hline\hline
 \mc{3}{c}{} & \mc{2}{c}{total}    & \mc{2}{c}{deuterium} &          \\ 
           &&& \tt cut0 & \tt cut3 &  \tt cut0 & \tt cut3 & CTEQ6.1  \\
           \hline\hline
  DIS      & JLab  &\cite{Malace}        & --  & 272 & --  & 136 &         \\
           & SLAC  &\cite{SLAC}          & 206 &1147 & 104 & 582 &         \\
           & NMC   &\cite{NMC}           & 324 & 464 & 123 & 189 & $\surd$ \\ 
           & BCDMS &\cite{BCDMS}         & 590 & 605 & 251 & 254 & $\surd$ \\
           & H1    &\cite{H1}            & 230 & 251 & --\ & --\ & $\surd$ \\
           & ZEUS  &\cite{ZEUS}          & 229 & 240 & --\ & --\ & $\surd$ \\
           \hline
 $\nu$A DIS & CCFR  &\cite{CCFR2,CCFR3}  & --\ & --\ & --\ & --\ & $\surd$ \\
           \hline
  DY       & E605 &\cite{E605}   &\mc{2}{c|}{119}&\mc{2}{c|}{--} & $\surd$ \\
           & E866 &\cite{E866}   &\mc{2}{c|}{375}&\mc{2}{c|}{191}&     \\
           \hline
$W$ asymmetry& CDF '98 ($\ell$) &\cite{CDF98}
                                 &\mc{2}{c|}{11} &\mc{2}{c|}{--} & $\surd$\\  
           & CDF '05 ($\ell$) &\cite{CDF05}
                                 &\mc{2}{c|}{11} &\mc{2}{c|}{--} &     \\  
           & D0 '08 ($\ell$)  &\cite{D008}
                                 &\mc{2}{c|}{10} &\mc{2}{c|}{--} &     \\  
           & D0 '08 ($e$)     &\cite{D0_e08}
                                 &\mc{2}{c|}{12} &\mc{2}{c|}{--} &     \\  
           & CDF '09 ($W$)    &\cite{CDF09}
                                 &\mc{2}{c|}{13} &\mc{2}{c|}{--} &     \\  
           \hline
jet        & CDF              &\cite{CDFjet}
                                 &\mc{2}{c|}{33} &\mc{2}{c|}{--} & $\surd$\\  
           & D0               &\cite{D0jet}
                                 &\mc{2}{c|}{90} &\mc{2}{c|}{--} & $\surd$\\  
           \hline
$\gamma$+jet& D0              &\cite{D0gamjet}
                                 &\mc{2}{c|}{56} &\mc{2}{c|}{--} &     \\  
\hline\hline
\multicolumn{3}{c|}{TOTAL} & 2408 & 3709 & 569 & 1161 &        \\
\hline\hline \\
\end{tabular}
\label{tab:datasets}
\end{table}

\begin{figure}
\includegraphics[width=0.48\linewidth,bb=20 165 570 700,clip=true]
                {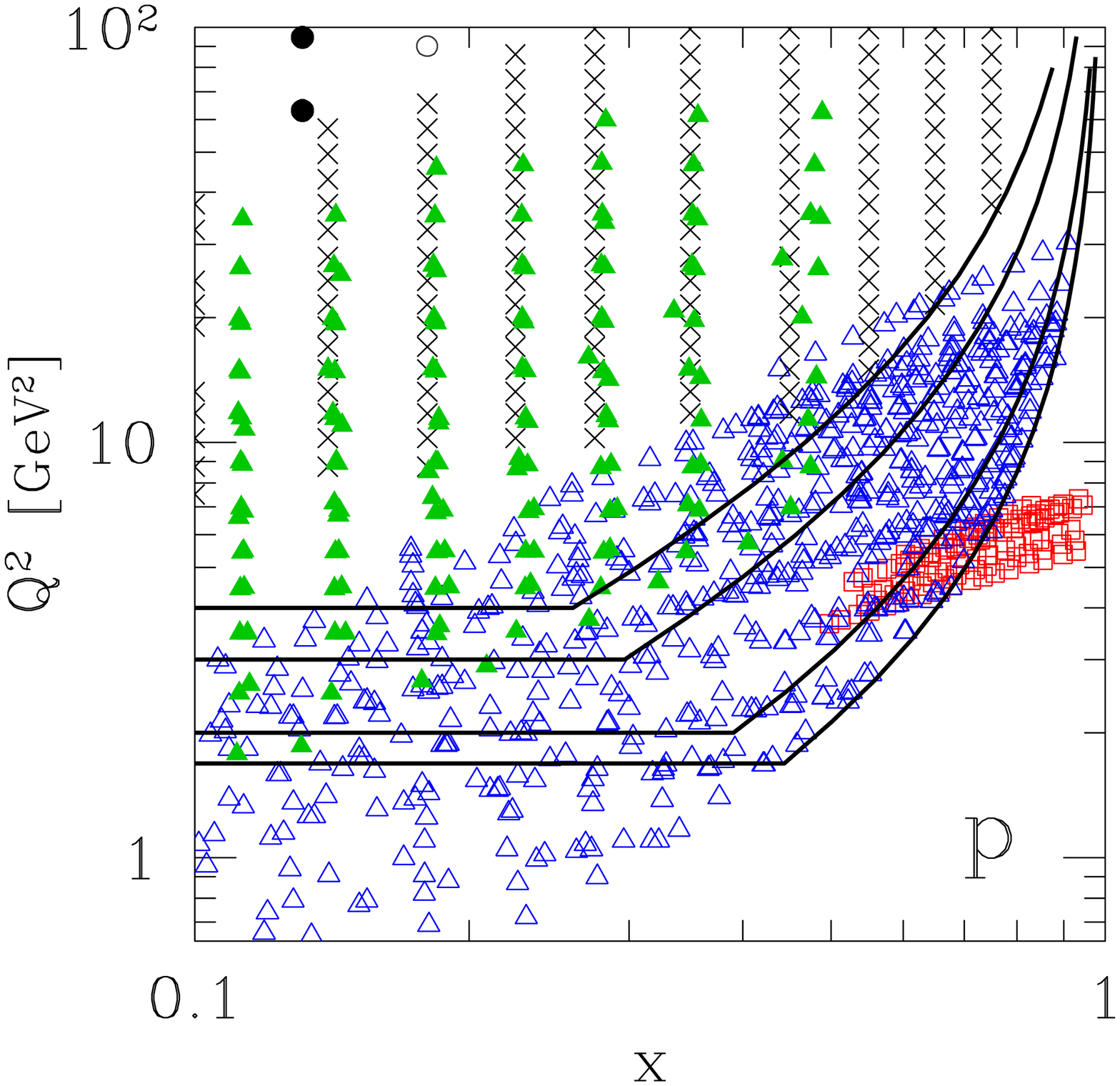} \  
\includegraphics[width=0.48\linewidth,bb=20 165 570 700,clip=true]
                {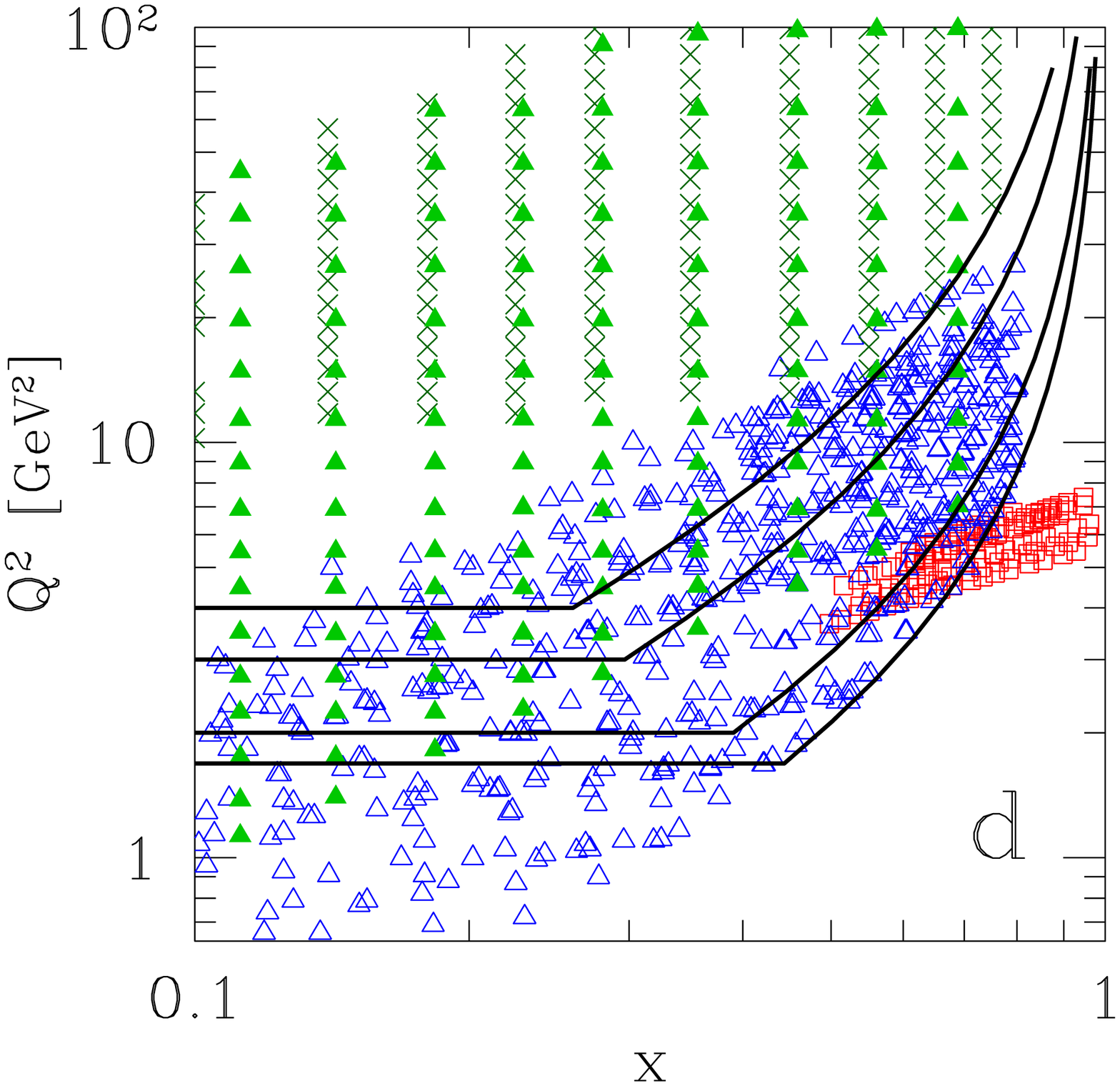} 
\caption{DIS data points and cuts for proton (left) and deuterium
	(right) targets, with the {\tt cut0}--{\tt cut3} kinematic
	cuts drawn from top to bottom with solid lines.
	The data points are from JLab (open red squares),
	SLAC (open blue triangles), NMC (filled green triangles),
	BCDMS (crosses), H1 (filled black circles),
	ZEUS (open black circles).}
\label{fig:DISps}
\end{figure}

The standard cuts on DIS data in previous global analyses have excluded
data with $Q^2 < 4$~GeV$^2$ and $W^2 < 12.25$~GeV$^2$, which means that
effectively the PDFs have only been constrained up to $x \approx 0.7$.
As discussed in Sec.~\ref{sec:intro}, extending the coverage to
larger $x$ requires relaxing the $Q^2$ and $W^2$ cuts.
To test the effects on the PDFs as the kinematic cuts are relaxed, 
we consider several cuts in addition to the standard one, which we
denote as follows:
\begin{itemize}
\item {\tt cut0}:  $Q^2 > 4$~GeV$^2$, $W^2 > 12.25$~GeV$^2$\ \ (standard)
\item {\tt cut1}:  $Q^2 > 3$~GeV$^2$, $W^2 > 8$~GeV$^2$
\item {\tt cut2}:  $Q^2 > 2$~GeV$^2$, $W^2 > 4$~GeV$^2$
\item {\tt cut3}:  $Q^2 > m_c^2 = 1.69$~GeV$^2$, $W^2 > 3$~GeV$^2$.
\end{itemize}
In Figure~\ref{fig:DISps} we plot the $(x,Q^2)$ coverage of the DIS
data sets used in this analysis, along with the above kinematic cuts.
The approximate doubling of the number of data points included in
the fits going from {\tt cut0} ($N_\text{DIS}=1579$) to {\tt cut3}
($N_\text{DIS}=3079$) is evident.
In particular we can include most of the SLAC data points, which reach
the lowest $Q^2$ values, and about half of the recent JLab data points.
At the lower $Q^2$ values, we include only those JLab data for which
an estimate of the uncertainties related to parametrizations of the 
longitudinal to transverse cross section ratio is available.

The methodology used in the global fits follows that of the CTEQ6
series of global fits \cite{CTEQ6.1}. 
In particular, the same functional forms for the PDF parametrizations
at the initial scale $Q_0=1.3$ GeV are employed.
Several of the parameters are fixed at representative values
and a total of 20 PDF parameters are varied in the fits.
An additional 3 free parameters are introduced for the
parametrization of the higher twist corrections, and error bands
are calculated using the Hessian technique. 
The $\Delta \chi$ tolerance used to define the errors in global
fits has been extensively discussed in the literature, and various 
groups use values ranging from $\Delta \chi$=1 to 10. 
We quote the error bands with $\Delta \chi$=1, and 
simply use them to show the relative variation of the
errors as the cuts are varied and between different PDFs.

%%%%%%%%%%%%%%%%%%%%%%%%%%%%%%%%%%%%%%%%%%%%%%%%%%%%%%%%%%%%%%%%%%%%%%%%%
\section{Fit results}
\label{sec:results}

Having described the expanded data sets used in this study and the new
ingredients necessary for fitting the data in the large-$x$ region,
in this section we present the numerical results of our analysis.

% .......................................................................
\subsection{Reference fit and kinematic cuts}

The first step in this study is to define a reference fit in order to
have a standard against which to compare the results of subsequent
fits as new terms are added and new sets of cuts are imposed.
This is achieved using the data listed in Table~I with the {\tt cut0}
set of kinematic cuts.
Target mass, higher twist and nuclear corrections (for deuterium data)
are {\it not} included in the reference fit, which should therefore
yield results similar to those of the CTEQ6.1 PDFs \cite{CTEQ6.1}.
Small differences can be expected, however, due to some additional
data sets used.
For example, the inclusion of the SLAC and E866 data induces
a small reduction of the $u$-quark distribution at large $x$.
The E866 data also lead to a sizable enhancement of the $d$-quark
distribution at large $x$, which is partially offset by the
$W$-asymmetry and $\gamma$+jet data.

\begin{figure}[tb]
\includegraphics[width=0.49\linewidth,bb=10 315 360 700,clip=true]
                {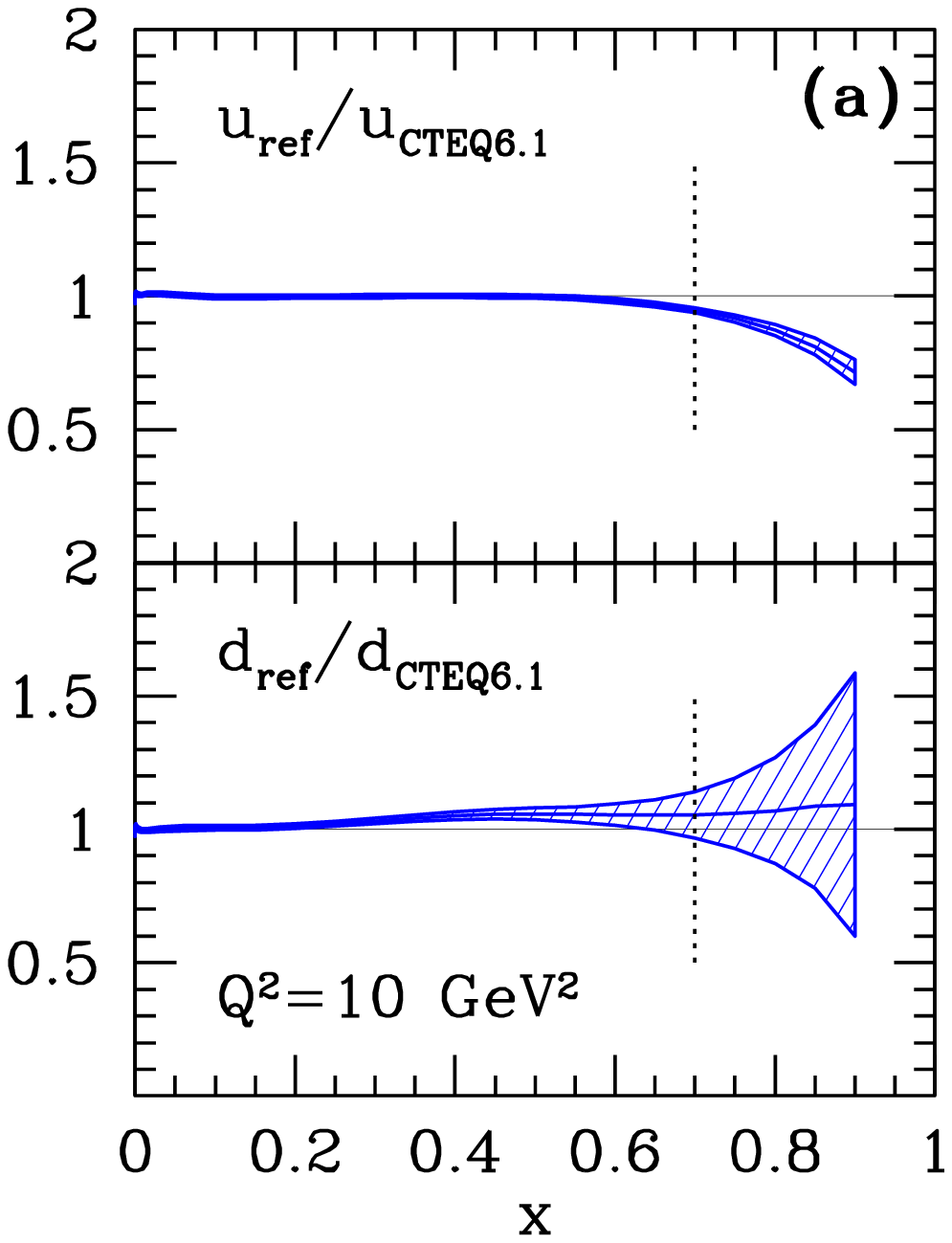}
\includegraphics[width=0.49\linewidth,bb=10 315 360 700,clip=true]
                {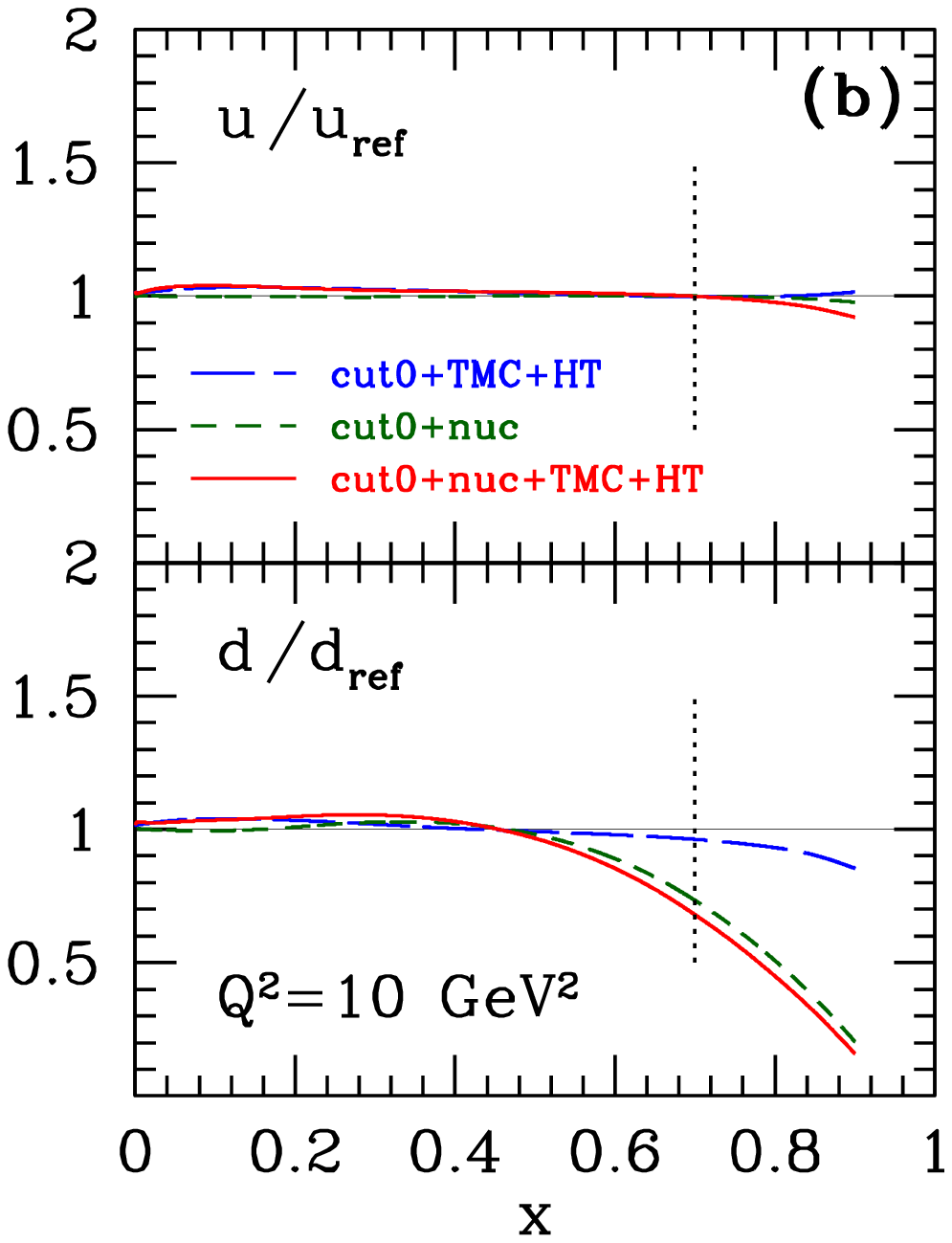}
\vskip-0.5cm
\caption{{\bf (a)}
	Impact of the expanded data set, with the standard cuts
	({\tt cut0}) and no target mass, higher twist or nuclear
	corrections applied.
	Illustrated are ratios of the PDFs from the resulting
	``reference'' fit to those from the CTEQ6.1 fit.
	{\bf (b)}
	Effects on the reference fit as TMC (using the CF prescription),
	higher twist (HT) and nuclear corrections (WBA) are added.}
\label{fig:newdata}
\end{figure}

The results for the $u$\ and $d$ distributions are shown in
Fig.~\ref{fig:newdata}(a) as ratios to the corresponding CTEQ6.1 PDFs. 
As anticipated, the $u$ distribution is slightly reduced in the
large-$x$ region, while the $d$ distribution shows a small increase
for $x \gtrsim 0.3$, leading to a modest increase of the $d/u$ ratio
at intermediate $x$.
The dotted vertical line at $x=0.7$ indicates the restriction on the
range in $x$ for which there are sufficient DIS data to constrain
the PDFs with {\tt cut0} kinematic limits.
Accordingly, the portion of each curve extending beyond the dotted
line represents an extrapolation of the fitted results.

The gluon and sea quark PDFs are not discussed here because they
are too poorly constrained at large $x$ by the chosen data sets.
More specifically, the gluon distribution is partially constrained
at intermediate and low $x$ by the scaling violations in DIS and
by the jet data.  At larger $x$, however, it is only indirectly
constrained by the momentum sum rule.
This can be seen by considering the effects of a small change in
one of the valence distributions, such as described above, which
alters their average momentum.
With the gluon already constrained somewhat in the intermediate-
to low-$x$ region, it is the large-$x$ gluon which is modified in
order to satisfy the momentum sum rule; since the gluon distribution
is much smaller than the valence distributions at large $x$, this
results in large variations of the high-$x$ gluons in order to 
compensate the rather small changes in the valence PDFs. 
The large variations in the shape of the gluon PDF at large $x$
then propagate to the sea PDFs via the actions of the evolution
equations.
Discussions on how to better experimentally constrain the gluon
distributions at large $x$ are contained in
Refs.~\cite{Alekhin03,Alekhin06}.

Having verified that the earlier CTEQ6.1 results could be reproduced
up to small variations due to the different data sets used, the next
step consists of a systematic investigation of the effects of
including target mass, higher twist, and nuclear corrections,
while also reducing the cuts on $W$\, and $Q$.
It is instructive to first consider the impact of each of these
corrections on the reference fit, which is illustrated in
Fig.~\ref{fig:newdata}(b). 
One can see that the {\tt cut0} constraints effectively limit any
change to the reference fit $u$ PDF.
On the other hand, while the $d$ PDF shows little sensitivity to the
target mass and higher twist effects, it does show that the nuclear 
corrections to the deuterium data have a profound effect.
The nuclear effects, as calculated in the framework of the weak
binding approximation \cite{WBA,KP,KMK}, cause a large reduction in
the $d$ PDF relative to the reference fit starting at $x \approx 0.5$.
The lesson for PDF global fits is that, even with the {\tt cut0}
constraints on the data selection, one should include nuclear
corrections for the deuterium data.
The {\tt cut0} constraints, however, effectively remove the need
for the inclusion of the TMC and HT contributions, as was the
original intent (which also confirms the analogous results of 
Ref.~\cite{Blumlein:2006be}). 
Of course, the PDFs are then directly constrained only for
$x \lesssim 0.7$.

\begin{figure}
\includegraphics[width=14cm,bb=40 380 570 700,clip=true]{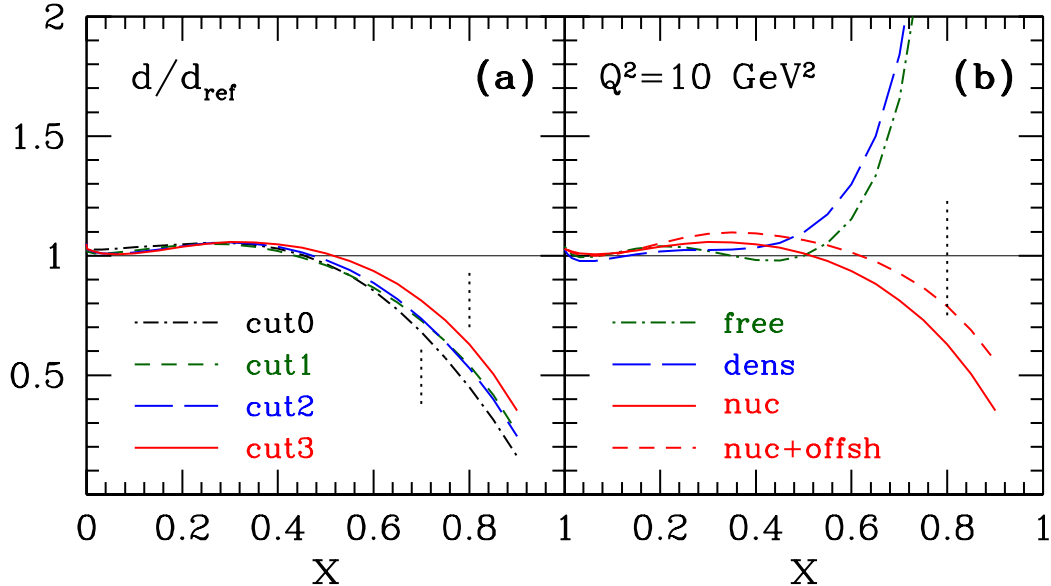}
\vspace*{-.5cm}
\caption{{\bf (a)}
	Effect on the $d$-quark distribution from varying the $Q$
	and $W$ cuts, including target mass (CF), higher twist and
	nuclear corrections (using the WBA nuclear smearing model).
	The vertical dotted lines indicate the $x$ region which is
	not directly constrained by DIS data ($x\gtrsim0.7$ for 
	\texttt{cut0} and $x\gtrsim0.8$ for \texttt{cut3}). 
	{\bf (b)}
	Sensitivity of the $d$-quark distribution to nuclear
	correction models with \texttt{cut3} kinematics:
	no nuclear effects (``free'', dot-dashed),
	nuclear density model (long-dashed),
	WBA nuclear smearing model without (solid) and with
	(short-dashed) off-shell corrections.}
\label{fig:nuc}
\end{figure}

With the reference fit as a starting point we next turn on the TMC,
add the higher twist parametrization, include the deuteron nuclear
correction, and then perform fits for the various sets of cuts
denoted by {\tt cut0} through {\tt cut3}. 
Progressively loosening the cuts naturally brings more data into play.
These data provide additional constraints at high values of $x$, and
one can see in Fig.~\ref{fig:nuc}(a) that the reduction in the $d$ PDF
is less pronounced than with {\tt cut0}.
Thus, the extrapolation of the {\tt cut0} results into the region
beyond $x\approx 0.7$ actually shows too strong a reduction.
Remarkably, the lowering of the $W$ and $Q$ cuts does not qualitatively
affect the $d$-quark suppression beyond $x \approx 0.5$, which remains
a stable feature for all fits considered.

% .......................................................................
\subsection{Sensitivity to nuclear corrections}

As suggested in Fig.~\ref{fig:newdata}(a) for {\tt cut3} kinematics,
and illustrated in Fig.~\ref{fig:nuc}(b) for {\tt cut3}, the behavior
of the $d$-quark distribution at large $x$ is driven primarily by the
nuclear effects.
With no nuclear corrections (treating the deuteron as a sum of a
free proton and free neutron), the $d$ quark is strongly enhanced
above $x \approx 0.5$.
This occurs because the Fermi motion in the deuteron causes an
enhancement of the data for $F_2^d$ beyond what would be expected
by treating the deuteron as the sum of free nucleons; if no smearing
corrections are included, the resulting $d$-quark distribution must
make up this enhancement as the $u$ distribution is already well
constrained at large values of $x$ by the proton data.
A qualitatively similar result is found using the nuclear density model
\cite{density}, which arises directly from the delayed onset of Fermi
motion in this model to $x \gtrsim 0.8$, and the corresponding smaller
deviation of the $F_2^d/F_2^N$ ratio from unity in this region.

On the other hand, correcting for the nuclear effects in the
$F_2^d$ data using the WBA finite-$Q^2$ smearing model discussed
in Sec.~\ref{ssec:nuclear} leads to a $d$-quark distribution with
the opposite trend at large $x$ relative to the reference fit,
as Fig.~\ref{fig:nuc}(b) illustrates.
At $x=0.8$ this amounts to a $\sim 40\%$ reduction of the $d$ PDF
relative to the reference fit.
Not accounting for the nuclear smearing in deuterium data will
thus lead to a significant overestimate of the $d$ distribution
at $x \gtrsim 0.6$.
This will be the case in fact for a wide range of nuclear smearing
models, and regardless of the details of the deuteron wave function.

Because the nucleons in the deuteron are slightly off-shell, with
$|p^2 - M^2|/M^2 = {\cal O}({\rm few}\ \%)$, in principle this can 
affect the extracted neutron structure function and resulting
$d$-quark distribution.
Using the model for off-shell corrections discussed in
Sec.~\ref{ssec:nuclear} does indeed result in an increase in the
$d$ distribution at large $x$ compared with the on-shell calculation,
which reflects the larger deviation from unity of the $F_2^d/F_2^N$
ratio in the presence of off-shell effects \cite{WBA,KP,MST1,MST2,GL}.
A simple way of understanding this is to consider the struck
nucleon as having a slightly smaller mass than a free nucleon,
$p^2 = M^{* 2} < M^2$.
In the target rest frame the effective scaling variable is then
$x^*=Q^2/2M^*\nu$, so the actual value of $x$ is larger than
would be expected for a free nucleon.
This leads to the observed structure function being slightly reduced
in this region, as can be seen from Eq.~(\ref{eq:delta_off}) where the
reduction is $\approx -1.5\%$.
%Applying this correction then leads to a slight increase in $F_2^d$.

To quantify the impact of the off-shell corrections on the $d$-quark
distribution one may consider the leading order structure functions
in the large-$x$ region, which is dominated by the valence PDFs,
\begin{equation}
F_2^p = \frac{4}{9}\, x\, u\, \left( 1 + \frac{d}{4u} \right) \ \ \ \ \
{\rm and} \ \ \ \ \
F_2^d = \frac{5}{9}\, x\, u\, \left( 1 + \frac{d}{u} \right)\, ,
\label{eq:F2LO}
\end{equation}
and examine the effect of a small variation in $F_2^d$ caused
by correcting for off-shell effects.
Since there is no variation in $F_2^p$, as an approximation the
proton structure function can be assumed to remain the same. 
The modifications to the $u$- and $d$-quark PDFs are then related
by $\delta u = -\frac{1}{4} \delta d$, and using Eq.~(\ref{eq:F2LO})
gives
\begin{equation}
\frac{\delta d}{d} = \frac{4}{3}\frac{\delta F_2^d}{F_2^d}
\left( 1+\frac{u}{d} \right)\, .
\end{equation}
A small change in $F_2^d$ thus results in a relative shift in the
$d$-quark distribution that is enhanced by a factor of order $u/d$.
For example, at $x=0.8$, the ratio $u/d \approx 20$, and a shift of 1.5\%
in $F_2^d$ leads to a increase of the $d$ distribution of about 40\%.
This is similar to what is observed in Fig.~\ref{fig:nuc}(b).

The same argument also explains the large sensitivity of the large-$x$
$d$-quark PDF to the choice of the nuclear correction model.
Currently this represents the single most important theoretical
uncertainty on large-$x$ PDFs, which can affect not only the
extraction of the $d/u$ ratio at large $x$ (see Sec.~\ref{sec:du}),
but also the determination of Tevatron and LHC parton luminosities
at large produced mass \cite{Alekhin01}.

\begin{figure}
\includegraphics[width=8cm,bb=18 180 592 718,clip=true]{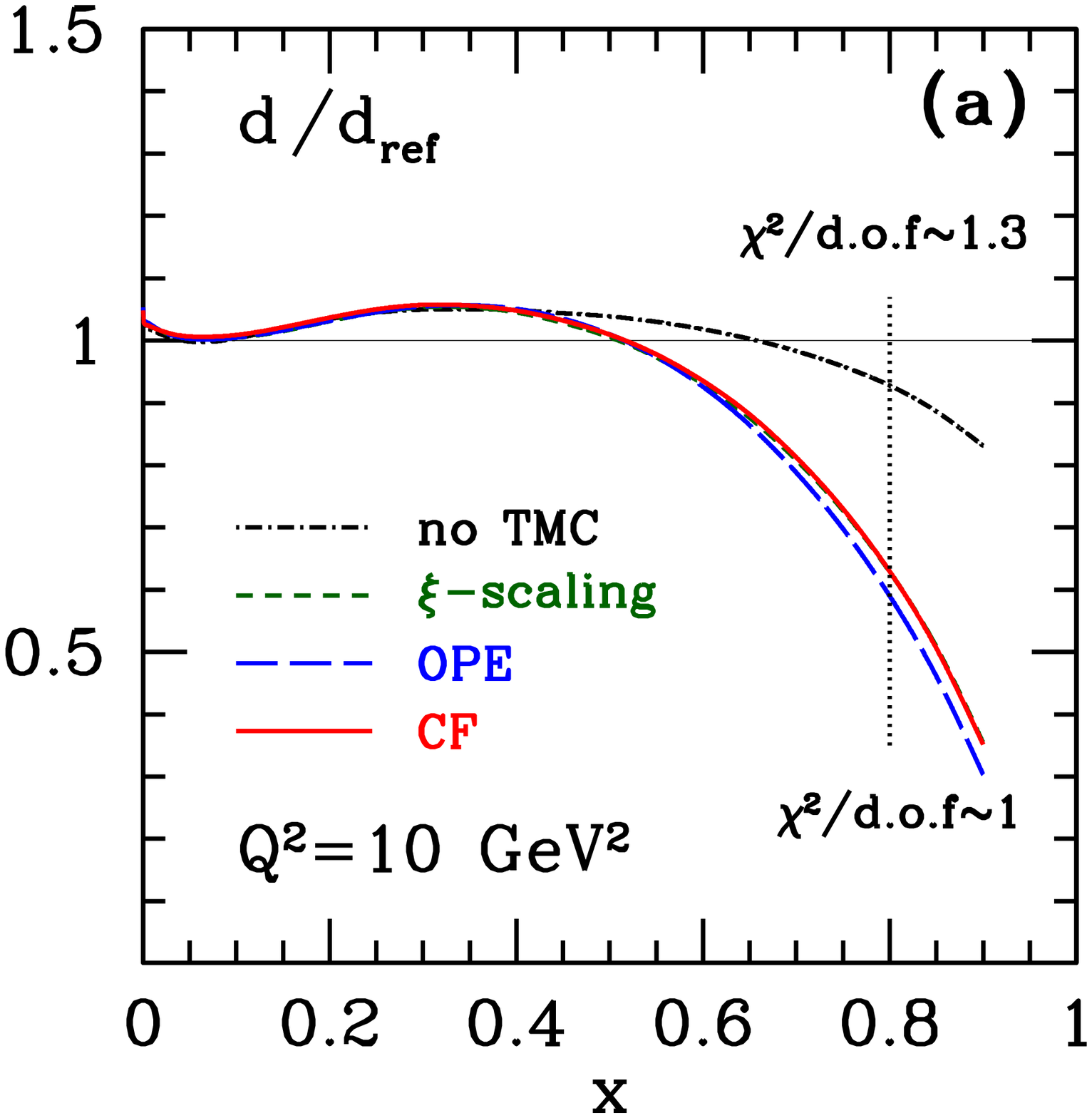}%
\includegraphics[width=8cm,bb=18 180 592 718,clip=true]{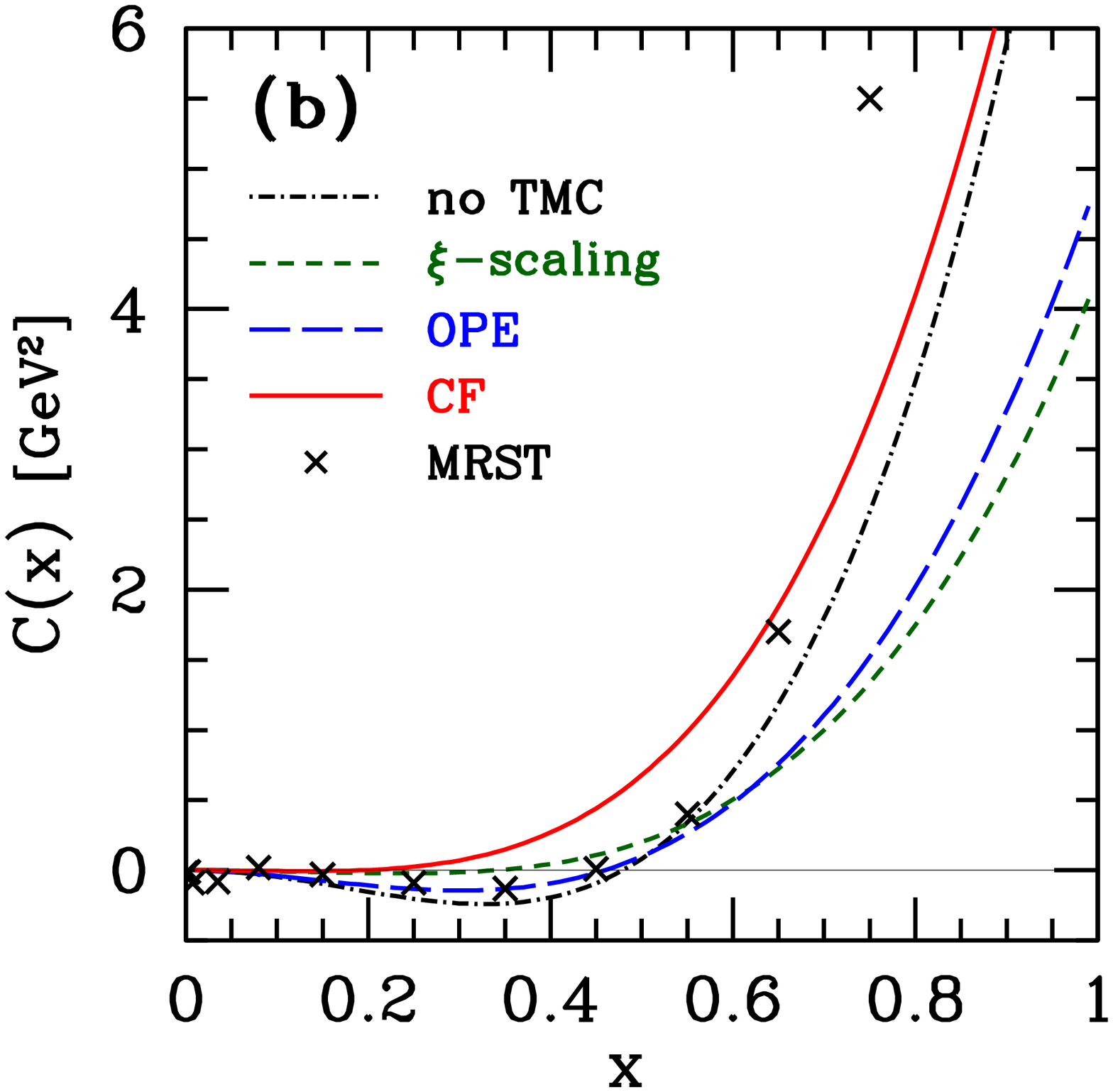}
\caption{{\bf (a)} 
        Effects of target mass and higher twist corrections on
	the $d$-quark distribution, using the $\xi$-scaling
	(short-dashed), OPE (long-dashed) and CF (solid) TMC
	prescriptions, as well as no TMC (dot-dashed).
	In each case a phenomenological higher twist correction is
	included in the fit, and the $\chi^2/d.o.f.$ is indicated.
	{\bf (b)}
	Extracted higher twist coefficient $C(x)$ for the different
	TMC prescriptions, compared with the higher twist
	determination from the MRST fit \cite{MRST_HT98}.}
\label{fig:HT}
\end{figure}

% .......................................................................
\subsection{Sensitivity to TMC and HT corrections}

The sensitivity of the $d$-quark distribution to the finite-$Q^2$
corrections associated with target mass and higher twist effects
is illustrated in Fig.~\ref{fig:HT}, where three different TMC
prescriptions are illustrated: the OPE-based prescription
\cite{GP,Schienbein}, the Nachtmann $\xi$-scaling
\cite{Aivazis,KretzerCF}, and the collinear factorization
prescription \cite{AQ}, along with a fit with no TMCs.
Although there is some dependence on the choice of TMCs, with the
addition of the phenomenological higher twist correction (\ref{eq:HT})
the resulting fits in Fig.~\ref{fig:HT}(a) are almost identical,
with $\chi^2/d.o.f. \sim 1$.
This reveals an important interplay between the target mass and higher
twist corrections, which tend to compensate each other in the fitting
procedure.

As discussed in Sec.~\ref{sec:theory}, given a sophisticated enough
parametrization of the higher twist corrections, one could in
principle accommodate the entire TMC effect, even though the
existence of TMCs is well-established theoretically.
In practice, however, it is preferable to include the TMC and HT
terms separately because of the different models in use for the TMCs,
and the separate HT parametrization is able to compensate the
different model choices.
Furthermore, it allows us to use a very economical functional
form for the higher twist corrections with only 3 parameters.
In contrast, with no TMCs we found no suitable parametrization
able to produce a satisfactory fit with a similar $\chi^2$.

Given that there is very little difference between the $d$-quark
distribution obtained using the different TMC choices, it is clear
that one cannot determine which is the preferred TMC prescription
from these fits alone.
On the other hand, the gluon distribution at large $x$ obtained by
analyzing the scaling violations of $F_2$ does depend somewhat on the 
chosen TMC prescription because of the different dependence on $Q^2$.
One can therefore include longitudinal structure function or cross 
section data in the fits to directly constrain the gluon distribution. 
A comparison of the fitted gluon distribution with the scaling 
violations of $F_2$ may then allow the TMC models to be distinguished.

The interplay between TMCs and higher twists is more vividly
demonstrated in Fig.~\ref{fig:HT}(b), where the higher twist
coefficient $C(x)$ is plotted for the various TMC prescriptions.
The characteristic rise with increasing $x$ is evident, with the higher
twist term being larger for the CF prescription than for the others.
The extracted higher twist coefficient for the case with no TMCs
is also qualitatively similar to that obtained in the MRST analysis
\cite{MRST_HT98}, which did not include target mass or deuterium
corrections.  The negative higher twist term at $x \sim 0.3$ evident
with the OPE and no TMC prescriptions disappears when using the CF
or $\xi$-scaling approaches.

The extracted HT coefficient for the OPE prescription is about
half of that obtained by Bl\"umlein {\it et al.} (BBG)
\cite{Blumlein:2006be,Blumlein:2008kz} in fixed-$x$ bins.
This may be due in part to our HT parametrization being too restrictive.
On the other hand, the BBG parton distributions have been fitted with
\texttt{cut0} kinematic cuts and then extrapolated to smaller $x$ and
$Q^2$ before extracting the HT terms.
This extrapolation may underestimate the $d$ quark distribution
at large $x$, similarly to our result in Figure~\ref{fig:nuc}(a),
therefore leading to an overestimate of the HT terms.

We emphasize, however, that our main finding here is not the
magnitude of the higher twist correction {\it per se}, but the
fact that very stable fits to {\em leading twist} PDFs for the $u$
and $d$ quarks can be obtained by the inclusion of both the target
mass and higher twist corrections.

% .......................................................................
\subsection{Final fit results}

\begin{figure}[tb]
\includegraphics[width=0.49\linewidth,bb=10 315 360 700,clip=true]
                {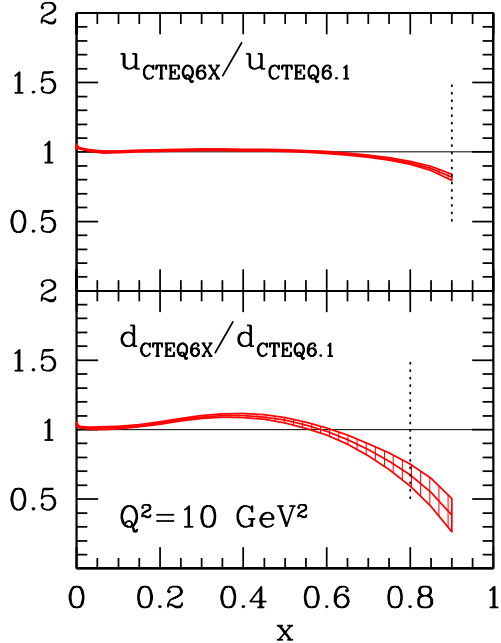}
\vskip-0.5cm
\caption{Results of the CTEQ6X fit with expanded kinematics
	({\tt cut3}) and inclusion of TMC, HT and nuclear corrections,
	normalized to the CTEQ6.1 PDFs.  The vertical lines show the
	approximate values of $x$ above which PDFs are not directly
	constrained by data.  The error bands correspond to
	$\Delta\chi=1$.}
\label{fig:cteqXcut3}
\end{figure}

We conclude this section with the presentation in
Fig.~\ref{fig:cteqXcut3} of the final fit results for the {\tt cut3}
set of constraints, which we refer to as ``CTEQ6X''.
One can see the significant reduction in the $d$ PDF at large values
of $x$ that was discussed previously, with the vertical lines again
indicating the approximate limits of the fit constraints.
As shown in Fig.~\ref{fig:DISps}, the proton DIS data extend to $x=0.9$,
and consequently the $u$ distribution is well constrained up to that
limit.  The $d$ distribution, however, is constrained only up to
$x=0.8$, reflecting the maximum extent of the deuterium data.
The error bands on the PDFs correspond to $\Delta\chi=1$ and include
systematic and statistical experimental uncertainties, but exclude
the theoretical uncertainties discussed above.

\begin{figure}[tb]
\includegraphics[width=12cm,bb=20 280 564 700,clip=true]{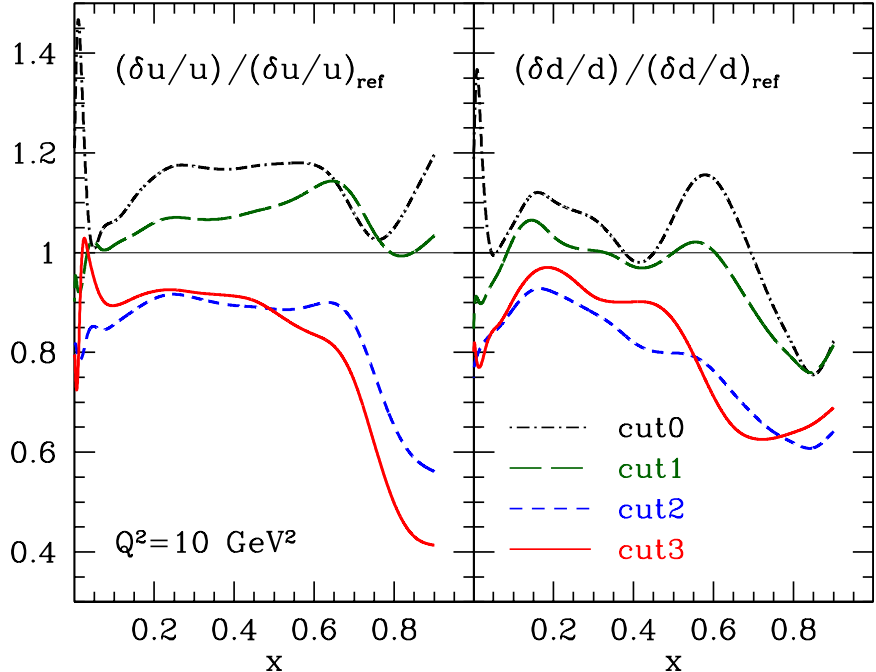}
\vskip-0.5cm
\caption{Relative PDF errors for $u$ and $d$ quarks normalized to
	the relative errors in the reference fit.}
\label{fig:relerr}
\end{figure}

One of the goals of this investigation was to examine the extent to
which the errors on the PDFs would be affected by the addition of the
new data made accessible by relaxing the selection criteria applied
to the DIS data.  This is shown in Fig.~\ref{fig:relerr} where the
relative errors on the $u {\rm \ and \ } d$ PDFs are shown, normalized
to the relative errors from the reference fit.  These ratios are shown
for the {\tt cut0} through {\tt cut3} kinematic cuts.  For \texttt{cut0}
and \texttt{cut1} the errors typically increase because the amount of
new DIS data is not sufficient to compensate for the added 3 free
parameters that describe the phenomenological higher twist
contributions.  For \texttt{cut2} and \texttt{cut3} there is a
substantial reduction in the uncertainty of these PDFs due to the
increased data, with the {\tt cut3} errors reduced by 10--20\% for
$x \lesssim 0.6$, and by up to 40--60\% at larger $x$.
However, it must be stressed that these errors only reflect
the uncertainties due to propagating the experimental errors.
In particular, uncertainties associated with nuclear corrections
are not shown; these will be discussed in more detail in a subsequent 
analysis \cite{future}.

\begin{figure}[tb]
\includegraphics[width=16.4cm,bb=30 160 565 390,clip=true]{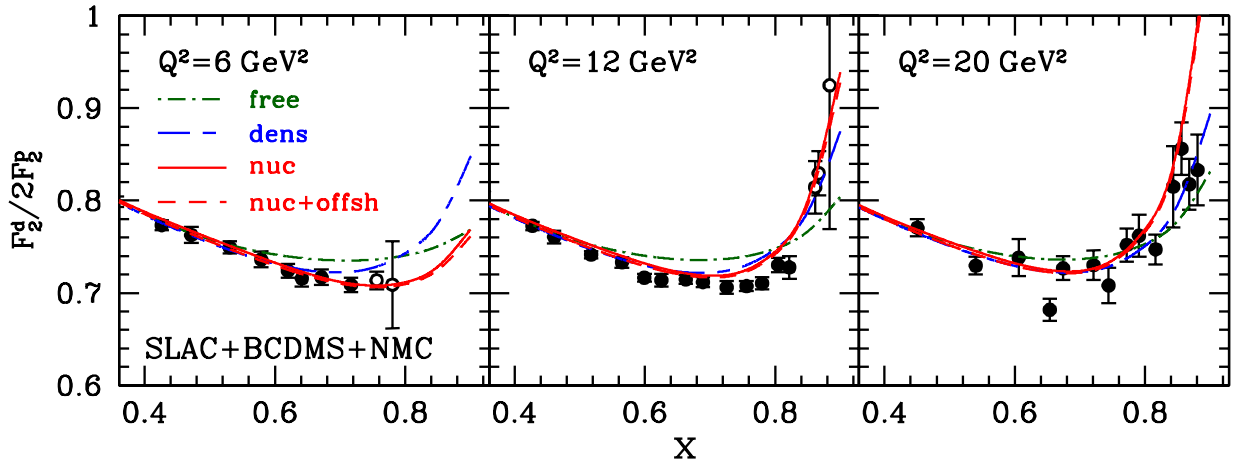}
\vskip-.5cm
\caption{Ratio of deuteron to proton $F_2$ structure functions
	computed from the \texttt{cut3} PDFs at $Q^2 = 6, 12$
	and 20~GeV$^2$ for various nuclear correction models,
	as in Fig.~\ref{fig:nuc}.  Data points are taken from the
	compilation of SLAC, BCDMS and NMC data in Ref.~\cite{Arrington},
	with $W^2<3$~GeV$^2$ data denoted by open circles.}
\label{fig:Dp}
\end{figure}

As a consistency check of the new fits, in Fig.~\ref{fig:Dp} we show
the ratio $F_2^d/2F_2^p$ computed from the new PDFs and compare them
to data at $Q^2 = 6, 12$ and 20~GeV$^2$ \cite{Arrington} using several
models for the nuclear corrections, as in Fig.~\ref{fig:nuc}(b).
The fit with no nuclear corrections is clearly disfavored, while
all three nuclear correction model curves show a $\chi^2$ which
is improved to some degree.
The nuclear density model correction has difficulty in reproducing
the $Q^2$ dependence of the data, especially at high $x$.
The WBA nuclear smearing model, on the other hand, is strongly
$Q^2$ dependent at large $x$ and rises more steeply than the density
model as $x \to 1$ when $Q^2 \gtrsim 10$~GeV$^2$.
The addition of off-shell corrections has very little effect on the fit:
essentially, the $d$-quark PDF shifts to compensate the inclusion of
the off-shell corrections and the PDFs obtained with these two models
provide an estimate of the uncertainty due to the model dependence of
the nuclear corrections.
A detailed comparison to data over a wider range of $Q^2$ would help to
clarify the nature of the $Q^2$ dependence of the nuclear corrections,
as also emphasized recently by Arrington {\it et al.} \cite{Arrington}.

%%%%%%%%%%%%%%%%%%%%%%%%%%%%%%%%%%%%%%%%%%%%%%%%%%%%%%%%%%%%%%%%%%%%%%%%%
\section{$\bm{d/u}$ ratio}
\label{sec:du}

According to spin-flavor SU(6) symmetry with no flavor-dependent
interactions between quarks, one would expect $d/u = 1/2$ for all $x$.
Empirically the $d/u$ ratio of course deviates strongly from this naive
expectation, and its $x \to 1$ behavior is a particularly sensitive
indicator of the dynamics responsible for the symmetry breaking.
If the interaction between the two valence quarks that are spectators
to the hard collision is mediated by a spin-dependent color-magnetic
force, such as from single gluon exchange (which also accounts for
the mass splitting between the nucleon and $\Delta$ \cite{DGG}),
the two-quark state with spin 0 will be energetically favored
relative to that with spin 1.
A dominant scalar ``diquark'' component of the proton would then
lead to a suppression of the $d$ quark distribution and a vanishing
$d/u$ ratio as $x \to 1$ \cite{quarter}.
On the other hand, if the dominant scattering process involves
scattering from quarks with the same helicity as that of the proton,
as would be expected from perturbative QCD, then this mechanism of
SU(6) breaking would lead to $d/u \to 1/5$ in the $x \to 1$ limit
\cite{FJ}.
The approach of the PDFs to the $x=1$ limiting values may also reveal
the role played by quark orbital angular momentum in the nucleon
\cite{Avakian}.

In this section we explore in more details what the new fits can tell
us about the large-$x$ behavior of the $d/u$ ratio.  We discuss the 
sensitivity of our $d/u$ fits to nuclear corrections, and review future
experiments which can more directly constrain the $d$ distribution
as $x \to 1$.

% .......................................................................
\subsection{Effect of nuclear corrections on $\bm{d/u}$}

\begin{figure}[tb]
\center
\includegraphics[height=7.5cm,bb=50 270 570 720,clip=true]{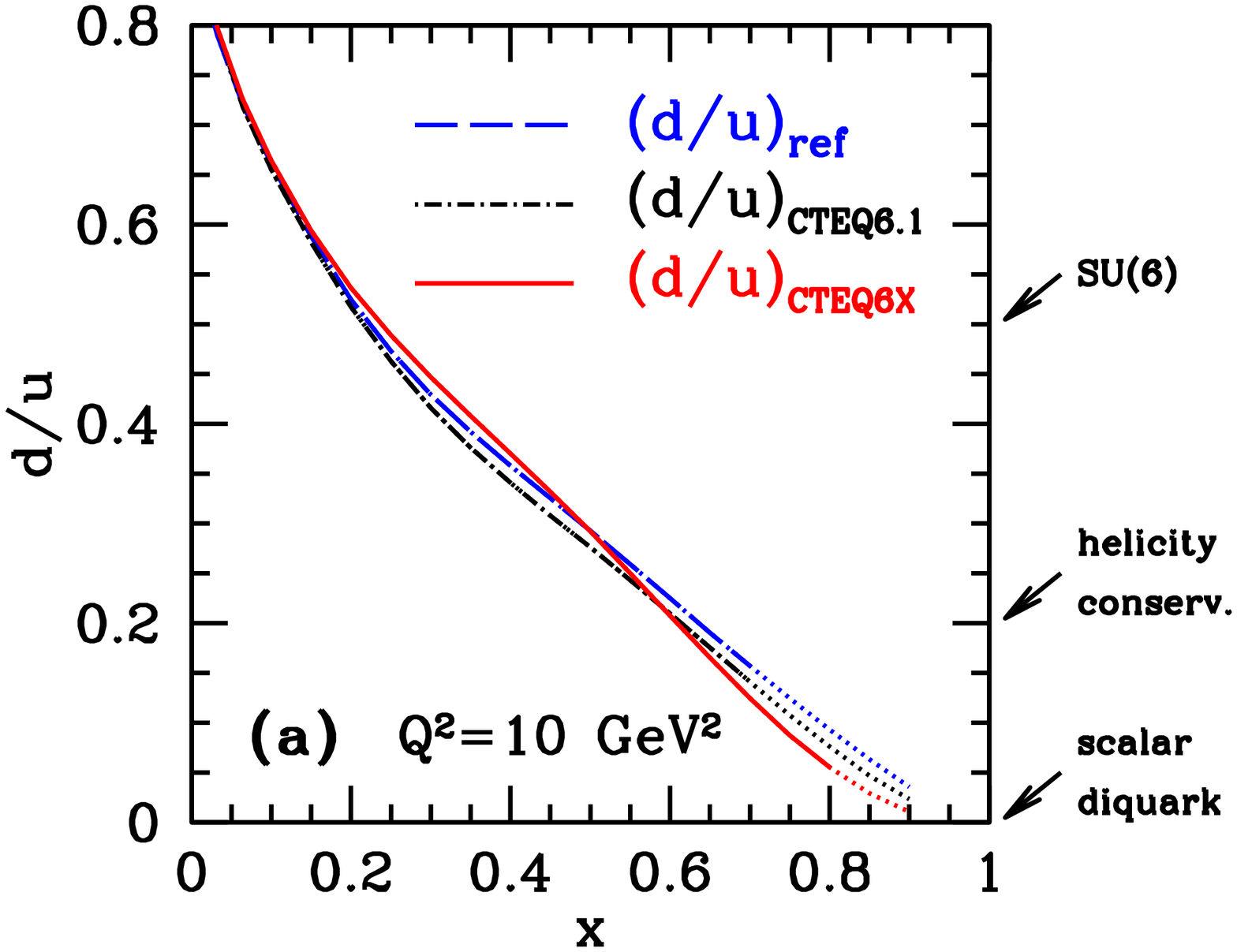}
\includegraphics[height=7.5cm,bb=50 270 470 720,clip=true]{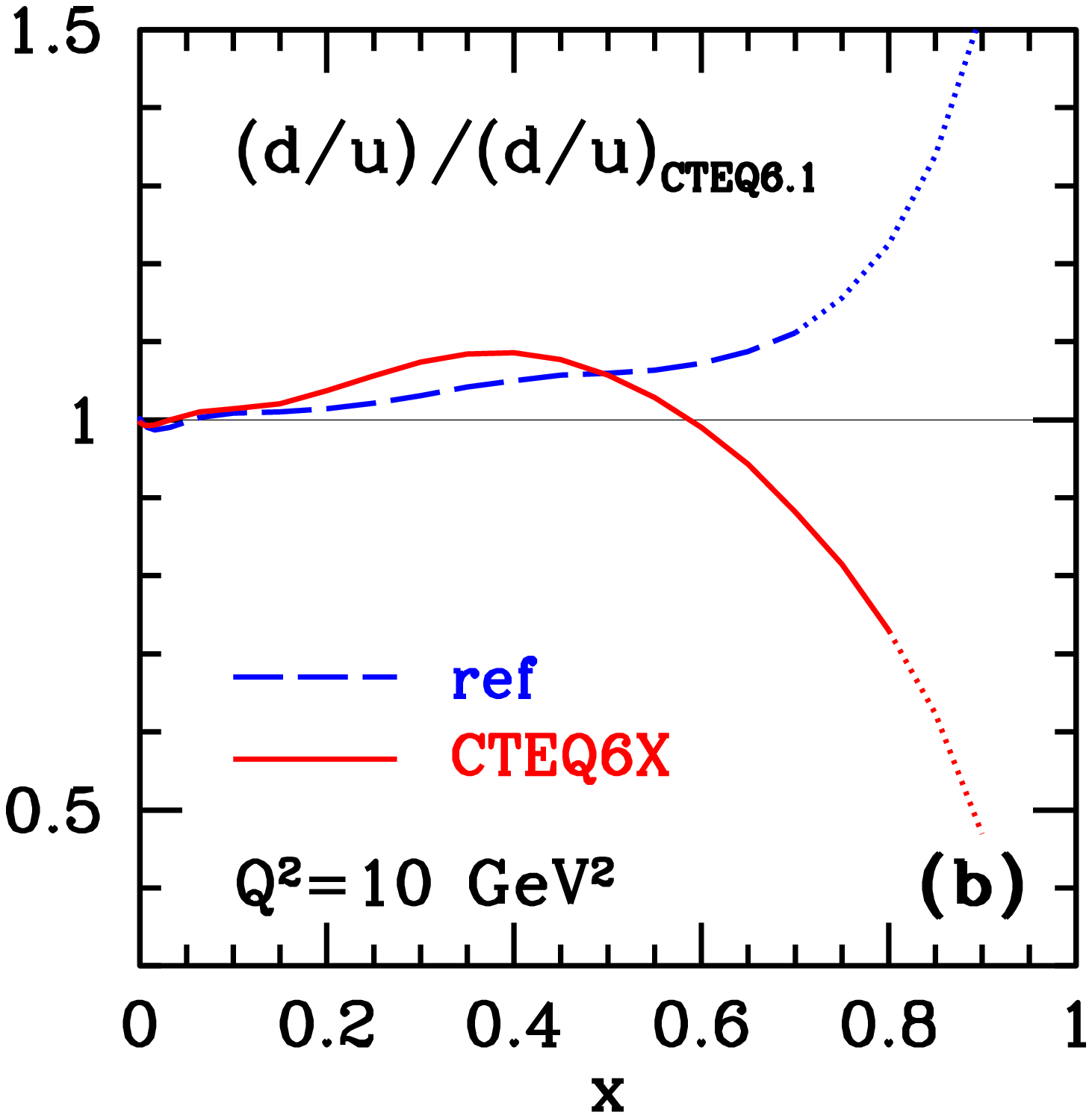}
\caption{{\bf (a)}
	$d/u$ quark distribution ratio for the reference, CTEQ6.1
	and CTEQ6X ({\tt cut3}) fits.
%to cut out the arrows and model names, use \texttt{bb=50 270 463 690}
	{\bf (b)}
	Ratio of the reference and CTEQ6X ({\tt cut3}) to the CTEQ6.1
	$d/u$ ratios. 
	In both panels the dotted lines indicate the region where
	the PDFs are not directly constrained by DIS data.}
\label{fig:du}
\end{figure}

The ratio of the $d$ to $u$ distributions is shown in Fig.~\ref{fig:du}
for the CTEQ6.1, reference, and CTEQ6X fits, together with predictions
from various models for the limiting behavior as $x \to 1$.
The rapid fall-off at high values of $x$ for the CTEQ6X PDFs
(with \texttt{cut3}) suggests that our results favor a lower $d/u$
value that is more consistent with the model of scalar diquark
dominance of the proton wave function than with models that predict
large $d/u$ asymptotic values.
In fact, relative to the standard PDF fits which assume no nuclear
corrections in the deuteron, the extracted $d/u$ ratio is reduced
at $x \gtrsim 0.6$, as evident already from Fig.~\ref{fig:nuc}.

We should note, however, that since our fitting region is restricted
to $x \lesssim 0.8$ for the $d$-quark distribution, the fits can only
extrapolate the behavior of the $d/u$ ratio at $0.8 \lesssim x \leq 1$.
Furthermore, since similar parametric forms are used for both the
$u$ and $d$ distributions, the $d/u$ ratio is constrained to approach
0 or infinity at the point $x=1$.
Nevertheless, in the absence of a dramatically large upturn in $d/u$
beyond $x \sim 0.8$, the fits in the constrained region suggest a
trend of the ratio towards a small limiting value.

This result is qualitatively different from the fit of Yang and Bodek
\cite{Yang}, in which the ratio of deuteron to proton and neutron $F_2$
structure functions was fitted with $d/u$ constrained to approach
the limit 1/5 at $x=1$ \cite{FJ}.
However, the neutron structure function in that analysis was extracted
assuming the nuclear density model \cite{density} for the nuclear
corrections in the deuteron.
As discussed in Sec.~\ref{ssec:nuclear}, the extrapolation of the
density model to the deuteron suffers from ambiguities associated
with defining physically meaningful nuclear densities for the deuteron
\cite{comment}, and even for large nuclei the model should not be
considered quantitative at very large $x$ ($x \gtrsim 0.6$--0.7)
\cite{density}.

In contrast to the nuclear density model, where the nuclear correction
is assumed to be independent of the bound nucleon structure, the size
of the nuclear correction within the nuclear smearing approach depends
also on the shape of the bound nucleon structure function through the
convolution in Eq.~(\ref{eq:F2d}).
Functions with harder $x$ distributions produce ratios $F_2^d/F_2^N$
which have greater depletion at intermediate and large $x$, and a
delayed onset of the Fermi motion rise as $x \to 1$. 
This was observed in the quark model calculation of Ref.~\cite{MST2},
which also illustrates the importance of the $Q^2$ dependence of
the data at large $x$.
Since the determination of the model parameters in \cite{MST2}
involved fitting to the large-$x$ proton and deuteron data without
including target mass and higher twist corrections, the resulting
distributions were generally harder than the empirical PDFs, leading
to a ratio $F_2^d/F_2^N$ in which the Fermi motion rise does not
appear until $x \gtrsim 0.8$.
Incorporating the finite-$Q^2$ corrections in the fits would
result in softer distributions and consequently an earlier rise
of $F_2^d/F_2^N$ due to Fermi motion at $x \sim 0.6$.

It is clear that obtaining a precise behavior of $d/u$ as $x \to 1$
requires a similarly precise knowledge of the nuclear corrections 
associated with the use of a deuterium target.
While the results will inevitably have some dependence on the nuclear
model adopted, our finding of a reduced $d$-quark distribution will
prevail in a wide range of models which incorporate the standard
nuclear corrections such as Fermi motion, binding and nucleon
off-shell effects.
As seen in Fig.~\ref{fig:Dp}, inclusive DIS data on a deuterium
target alone have little discriminatory power, except if the
$Q^2$ range and precision of the measurements are increased. 
A more detailed analysis of the correlation between nuclear
corrections and the $d/u$ ratio at $x \to 1$, as well as of the 
dependence of $d/u$ on the parametric forms, will be presented
in a future publication \cite{future}.

% .......................................................................
\subsection{Direct experimental constraints on $\bm{d/u}$}

To avoid nuclear corrections altogether, one would ideally need data
on free nucleon targets which would constrain the $d$-quark PDF.
This suggests weak interaction processes where one can utilize 
variations of the flavor changing transition $W^+ d \to u$. 
An example would be the DIS processes $\nu p \to \mu^- X$
and $\bar\nu p \to \mu^+ X$, which could probe the $d$ and $u$
PDFs, respectively, at large values of $x$.
However, existing data sets for neutrino scattering from hydrogen
do not have sufficient statistics to provide information in the
$x \to 1$ region.
A new high-statistics experiment on hydrogen, as is being considered
by the MINERvA collaboration \cite{MINERvA}, would be required for
such a study.

A further possibility is provided by the $W$-lepton or $W$ charge
asymmetries measured at the Tevatron in $p\bar p$ collisions.
These are sensitive to the $d/u$ ratio, and data at large values
of the lepton or $W$ rapidity can reach $x$ values as high as 0.8.
However, this is at the edge of the kinematic coverage and the
data there are statistically limited.
Nevertheless, new data sets with increased statistics may help 
provide additional constraints on $d/u$ which are independent
of nuclear corrections.

Another method which utilizes the weak interactions to probe the
$d$ quark involves parity-violating electron DIS on a hydrogen
target \cite{Souder,Hobbs}.
Here the asymmetry between left- and right-hand polarized electrons
selects the interference between the $\gamma$ and $Z$-boson exchange,
which depends on the $d/u$ ratio weighted by electroweak charges.
Such an experiment is planned at Jefferson Lab, taking advantage
of the high luminosity and energy available following the 12~GeV
energy upgrade, and with the expected 1\% measurements of the
asymmetry it would strongly constrain the $d/u$ ratio up to
$x \sim 0.8$ \cite{Souder}.

While still relying on nuclear targets, a novel idea proposes the
use of mirror symmetric nuclei, such as $^3$He and $^3$H, in which
the nuclear effects mostly cancel.
Explicit calculations have confirmed that accurate extraction of
$F_2^n/F_2^p$ is possible from measured ratios of $^3$He to $^3$H
structure functions, with nuclear corrections canceling to within
$\sim 1\%$ up to $x \approx 0.85$ \cite{A3}.

Finally, an experimental program is under way to determine the
$F_2^n$ structure function from measurements of DIS on a deuterium
target with low-momentum spectator protons in the backward
center-of-mass hemisphere, which tags DIS on an almost free
neutron in the deuteron \cite{BONUS}.
Preliminary results have confirmed the feasibility of this method,
and future measurements of spectator protons at the 12~GeV energy
updated Jefferson Lab, together with $^3$He or $^3$H spectators on
$^4$He targets, will be helpful both in determining the $d$ quark
at large $x$ and in constraining the nuclear correction models.

%%%%%%%%%%%%%%%%%%%%%%%%%%%%%%%%%%%%%%%%%%%%%%%%%%%%%%%%%%%%%%%%%%%%%%%%%
\section{Conclusion}
\label{sec:conc}

In this analysis we have explored the possibility of extending the
range of fitted DIS data in both $W$\ and $Q$ to lower values than
have been traditionally used in global PDF fits, in order to obtain
an extension of the covered range of $x$.  The results presented here
show that excellent fits can result from such a procedure provided
that target mass corrections, higher twist contributions, and nuclear
corrections for deuterium targets are all taken into account.
The resulting fits show that the leading twist $u$- and $d$-quark
PDFs are stable with respect to the choices made for implementing
the target mass corrections, as long as a flexible higher twist
parametrization is employed.

The major new feature of our fits compared to previous global analyses
is the stronger suppression of the $d$-quark distribution at large $x$.
However, the precise amount of suppression is sensitive to the treatment
of the nuclear corrections, which are an important source of theoretical
uncertainty at $x \sim 1$.  Hence, further progress in the determination
of the behavior of the large-$x$ PDFs and the $d/u$ ratio requires
either a better understanding of the nuclear corrections or the use of
data obtained using free nucleons in the initial state, for which we
reviewed several experimental possibilities.

%%%%%%%%%%%%%%%%%%%%%%%%%%%%%%%%%%%%%%%%%%%%%%%%%%%%%%%%%%%%%%%%%%%%%%%
\begin{acknowledgments}

We thank J.~Arrington and S.~P.~Malace for helpful communications.
This work has been supported by the DOE contract No. DE-AC05-06OR23177,
under which Jefferson Science Associates, LLC operates Jefferson Lab,
DOE grant No. DE-FG02-97ER41022, and NSF award No.~0653508.

\end{acknowledgments}

%%%%%%%%%%%%%%%%%%%%%%%%%%%%%%%%%%%%%%%%%%%%%%%%%%%%%%%%%%%%%%%%%%%%%%%


\begin{thebibliography}{99}

\bibitem{Nadolsky}
P.~M.~Nadolsky {\it et al.},
Phys.\ Rev.\  D {\bf 78}, 013004 (2008).
%%CITATION = PHRVA,D78,013004;%%

\bibitem{MSTW}
A.~D.~Martin, W.~J.~Stirling, R.~S.~Thorne and G.~Watt,
arXiv:0901.0002 [hep-ph].
%%CITATION = ARXIV:0901.0002;%%

\bibitem{ABKM}
S.~Alekhin, J.~Bl\"umlein, S.~Klein and S.~Moch,
arXiv:0908.2766 [hep-ph].
%%CITATION = ARXIV:0908.2766;%%

\bibitem{Ball:2009mk}
  R.~D.~Ball {\it et al.}  [NNPDF Collaboration],
  %``Precision determination of electroweak parameters and the strange content
  %of the proton from neutrino deep-inelastic scattering,''
  Nucl.\ Phys.\  B {\bf 823}, 195 (2009).
  %[arXiv:0906.1958 [hep-ph]].
  %%CITATION = NUPHA,B823,195;%%

\bibitem{MEK}
W.~Melnitchouk, R.~Ent and C.~Keppel,
Phys.\ Rept.\  {\bf 406}, 127 (2005).
%%CITATION = PRPLC,406,127;%%

\bibitem{pQCD}
R.~Blankenbecler and S.~J.~Brodsky,
Phys.\ Rev.\  D {\bf 10}, 2973 (1974);
%%CITATION = PHRVA,D10,2973;%%
%
J.~F.~Gunion,
Phys.\ Rev.\  D {\bf 10}, 242 (1974);
%%CITATION = PHRVA,D10,242;%%
%
S.~J.~Brodsky and G.~P.~Lepage,
Proceedings of the 1979 Summer Institute on Particle Physics, SLAC (1979).
%``Perturbative Quantum Chromodynamics,''
%%CITATION = SLAC-PUB-2447;%%

\bibitem{MT_NP}
W.~Melnitchouk and A.~W.~Thomas,
Phys.\ Lett.\  B {\bf 377}, 11 (1996).
%%CITATION = PHLTA,B377,11;%%

\bibitem{Kuhlmann}
S.~Kuhlmann {\it et al.},
Phys.\ Lett.\  B {\bf 476}, 291 (2000).
%%CITATION = PHLTA,B476,291;%%

\bibitem{Itow:2001ee}
Y.~Itow {\it et al.}  [T2K Collaboration],
%``The JHF-Kamioka neutrino project,''
arXiv:hep-ex/0106019.
%%CITATION = HEP-EX/0106019;%%

\bibitem{Ayres:2004js}
D.~S.~Ayres {\it et al.}  [NOvA Collaboration],
%``NOvA proposal to build a 30-kiloton off-axis detector to study neutrino
%oscillations in the Fermilab NuMI beamline,''
arXiv:hep-ex/0503053.
%%CITATION = HEP-EX/0503053;%%

\bibitem{Raby:2008pd}
S.~Raby {\it et al.},
%``DUSEL Theory White Paper,''
arXiv:0810.4551 [hep-ph].
%%CITATION = ARXIV:0810.4551;%%

\bibitem{Alekhin01}
S.~Alekhin,
Phys.\ Rev.\  D {\bf 63}, 094022 (2001).
%%CITATION = PHRVA,D63,094022;%%

\bibitem{Alekhin03}
S.~Alekhin,
Phys.\ Rev.\  D {\bf 68}, 014002 (2003).
%%CITATION = PHRVA,D68,014002;%%

\bibitem{Alekhin06}
S.~Alekhin, K.~Melnikov and F.~Petriello,
Phys.\ Rev.\  D {\bf 74}, 054033 (2006).
%%CITATION = PHRVA,D74,054033;%%

\bibitem{CTEQ6.1}
J.~Pumplin, D.~R.~Stump, J.~Huston, H.~L.~Lai, P.~M.~Nadolsky and
W.~K.~Tung,
JHEP {\bf 0207}, 012 (2002); 
%%CITATION = JHEPA,0207,012;%%
%
D.~Stump, J.~Huston, J.~Pumplin, W.~K.~Tung, H.~L.~Lai, S.~Kuhlmann and
J.~F.~Owens,
JHEP {\bf 0310}, 046 (2003).
%%CITATION = JHEPA,0310,046;%%

\bibitem{Pumplin:2009sc}
J.~Pumplin,
%``Experimental consistency in parton distribution fitting,''
arXiv:0909.0268 [hep-ph].
%%CITATION = ARXIV:0909.0268;%%

\bibitem{Martin:2003sk}
A.~D.~Martin, R.~G.~Roberts, W.~J.~Stirling and R.~S.~Thorne,
Eur.\ Phys.\ J.\  C {\bf 35}, 325 (2004).
%%CITATION = EPHJA,C35,325;%%

\bibitem{West}
G.~B.~West,
Phys.\ Lett.\  B {\bf 37}, 509 (1971).
%%CITATION = PHLTA,B37,509;%%

\bibitem{Jaffe}
R.~L.~Jaffe,
in {\it Relativistic Dynamics and Quark-Nuclear Physics}, 
M.~B.~Johnson and A.~Pickleseimer editors
(Wiley, New York, 1985).

\bibitem{Bickerstaff}
R.~P.~Bickerstaff and A.~W.~Thomas,
J.\ Phys.\ G {\bf 15}, 1523 (1989).
%%CITATION = JPHGB,G15,1523;%%

\bibitem{GST}
D.~F.~Geesaman, K.~Saito and A.~W.~Thomas,
Ann.\ Rev.\ Nucl.\ Part.\ Sci.\  {\bf 45}, 337 (1995).
%%CITATION = ARNUA,45,337;%%

\bibitem{WBA}
S.~A.~Kulagin, G.~Piller and W.~Weise,
Phys.\ Rev.\  C {\bf 50}, 1154 (1994).
%%CITATION = PHRVA,C50,1154;%%

\bibitem{WBApol}
For applications of the WBA to spin dependent structure functions
see:
%
S.~A.~Kulagin, W.~Melnitchouk, G.~Piller and W.~Weise,
Phys.\ Rev.\  C {\bf 52}, 932 (1995);
%%CITATION = PHRVA,C52,932;%%
%
S.~A.~Kulagin and W.~Melnitchouk,
Phys.\ Rev.\  C {\bf 77}, 015210 (2008);
%%CITATION = PHRVA,C77,015210;%%
%
{\it ibid} C {\bf 78}, 065203 (2008).
%%CITATION = PHRVA,C78,065203;%%

\bibitem{KP}
S.~A.~Kulagin and R.~Petti,  
Nucl.\ Phys.\  A {\bf 765}, 126 (2006).
%%CITATION = NUPHA,A765,126;%%

\bibitem{KMK}
Y.~Kahn, W.~Melnitchouk and S.~A.~Kulagin,
Phys.\ Rev.\  C {\bf 79}, 035205 (2009).
%%CITATION = PHRVA,C79,035205;%%

\bibitem{AQV}
A.~Accardi, J.~W.~Qiu, J.~P.~Vary,
``Collinear factorization and deep inelastic scattering on nuclear
targets'', {\it in preparation}.

\bibitem{Paris}
M.~Lacombe {\it et al.},
% B.~Loiseau, J.~M.~Richard, R.~Vinh Mau, J.~Cote, P.~Pires, R.~De Tourreil,
Phys.\ Rev.\  C {\bf 21}, 861 (1980).
%%CITATION = PHRVA,C21,861;%%

\bibitem{AV18}
R.~B.~Wiringa, V.~G.~J.~Stoks and R.~Schiavilla,
Phys.\ Rev.\  C {\bf 51}, 38 (1995).
%%CITATION = PHRVA,C51,38;%%

\bibitem{CD-Bonn}
R.~Machleidt,
Phys.\ Rev.\  C {\bf 63}, 024001 (2001).
%%CITATION = PHRVA,C63,024001;%%

\bibitem{MST2}
W.~Melnitchouk, A.~W.~Schreiber and A.~W.~Thomas,
Phys.\ Lett.\  B {\bf 335}, 11 (1994).
%%CITATION = PHLTA,B335,11;%%

\bibitem{MST1}
W.~Melnitchouk, A.~W.~Schreiber and A.~W.~Thomas,
Phys.\ Rev.\  D {\bf 49}, 1183 (1994).
%%CITATION = PHRVA,D49,1183;%%

\bibitem{Gross}
W.~W.~Buck and F.~Gross,
Phys.\ Rev.\  D {\bf 20}, 2361 (1979);
%%CITATION = PHRVA,D20,2361;%%
%
F.~Gross and A.~Stadler,
Phys.\ Rev.\  C {\bf 78}, 014005 (2008).
%%CITATION = PHRVA,C78,014005;%%

\bibitem{Blumlein:2006be}
J.~Bl\"umlein, H.~B\"ottcher and A.~Guffanti,
Nucl.\ Phys.\  B {\bf 774}, 182 (2007).
%%CITATION = NUPHA,B774,182;%%

\bibitem{Yang}
U.~K.~Yang and A.~Bodek,
Phys.\ Rev.\ Lett.\  {\bf 82}, 2467 (1999).
%%CITATION = PRLTA,82,2467;%%

\bibitem{Gomez}
J.~Gomez {\it et al.},
Phys.\ Rev.\ D{\bf 49}, 4348 (1994).

\bibitem{density}
L.~L.~Frankfurt and M.~I.~Strikman,
Nucl.\ Phys.\  B {\bf 250}, 143 (1985);
%%CITATION = NUPHA,B250,143;%%
%
Phys.\ Rept.\  {\bf 160}, 235 (1988).
%%CITATION = PRPLC,160,235;%%

\bibitem{comment}
W.~Melnitchouk, I.~R.~Afnan, F.~R.~P.~Bissey and A.~W.~Thomas,
Phys.\ Rev.\ Lett.\  {\bf 84}, 5455 (2000).
%%CITATION = PRLTA,84,5455;%%

\bibitem{Nachtmann}
O.~Nachtmann,
Nucl.\ Phys.\  B {\bf 63}, 237 (1973).
%%CITATION = NUPHA,B63,237;%%

\bibitem{GP}
H.~Georgi and H.~D.~Politzer,
Phys.\ Rev.\  D {\bf 14}, 1829 (1976).
%%CITATION = PHRVA,D14,1829;%%

\bibitem{Tung}
K.~Bitar, P.~W.~Johnson and W.~k.~Tung,
Phys.\ Lett.\ B {\bf 83}, 114 (1979);
%
P.~W.~Johnson and W.~K.~Tung,
Print-79-1018 (Illinois Tech)
{\it Neutrino '79}, Bergen, Norway (1979).

\bibitem{Steffens}
F.~M.~Steffens and W.~Melnitchouk,
Phys.\ Rev.\  C {\bf 73}, 055202 (2006).
%%CITATION = PHRVA,C73,055202;%%

\bibitem{Schienbein}
I.~Schienbein {\it et al.},
J.\ Phys.\ G {\bf 35}, 053101 (2008).
%%CITATION = JPHGB,G35,053101;%%

\bibitem{EFP}
R.~K.~Ellis, W.~Furmanski and R.~Petronzio,
Nucl.\ Phys.\  B {\bf 212}, 29 (1983).
%%CITATION = NUPHA,B212,29;%%

\bibitem{Collins}
J.~C.~Collins, D.~E.~Soper and G.~Sterman,
Adv.\ Ser.\ Direct.\ High Energy Phys.\  {\bf 5}, 1 (1988).
%%CITATION = HEP-PH 0409313;%%

\bibitem{AQ}
A.~Accardi and J.~W.~Qiu,
JHEP {\bf 07}, 090 (2008).
%%CITATION = ARXIV:0805.1496;%%

\bibitem{Aivazis}
M.~A.~G.~Aivazis, F.~I.~Olness and W.~K.~Tung,
Phys.\ Rev.\  D {\bf 50}, 3085 (1994).
%%CITATION = PHRVA,D50,3085;%%

\bibitem{KretzerCF}
S.~Kretzer and M.~H.~Reno,
Phys.\ Rev.\  D {\bf 66}, 113007 (2002).
%%CITATION = PHRVA,D66,113007;%%

\bibitem{AM}
For an extension to spin dependent scattering see
A.~Accardi and W.~Melnitchouk,
Phys.\ Lett.\  B {\bf 670}, 114 (2008).
%%CITATION = PHLTA,B670,114;%%

\bibitem{AKP07}
S.~Alekhin, S.~A.~Kulagin and R.~Petti,
AIP Conf.\ Proc.\  {\bf 967}, 215 (2007).
%%CITATION = APCPC,967,215;%%

\bibitem{Pumplin:2002vw}
J.~Pumplin, D.~R.~Stump, J.~Huston, H.~L.~Lai, P.~M.~Nadolsky and
W.~K.~Tung,
JHEP {\bf 0207}, 012 (2002).
%%CITATION = JHEPA,0207,012;%%

\bibitem{Virchaux}
M.~Virchaux and A.~Milsztajn,
Phys.\ Lett.\  B {\bf 274}, 221 (1992).
%%CITATION = PHLTA,B274,221;%%

\bibitem{MRST_HT98}
A.~D.~Martin, R.~G.~Roberts, W.~J.~Stirling and R.~S.~Thorne,
Phys.\ Lett.\  B {\bf 443}, 301 (1998).
%%CITATION = PHLTA,B443,301;%%

\bibitem{Blumlein:2008kz}
J.~Bl\"umlein and H.~B\"ottcher,
Phys.\ Lett.\  B {\bf 662}, 336 (2008).
%%CITATION = PHLTA,B662,336;%%

\bibitem{Schaefer:2001uh}
S.~Schaefer, A.~Schafer and M.~Stratmann,
Phys.\ Lett.\  B {\bf 514}, 284 (2001).
%%CITATION = PHLTA,B514,284;%%

\bibitem{Alekhin:1999kt}
S.~Alekhin,
Phys.\ Lett.\  B {\bf 488}, 187 (2000).
%%CITATION = PHLTA,B488,187;%%

\bibitem{Qiu:1988dn}
J.~W.~Qiu,
Phys.\ Rev.\  D {\bf 42}, 30 (1990).
%%CITATION = PHRVA,D42,30;%%

\bibitem{AKL04}
S.~Alekhin, S.~A.~Kulagin and S.~Liuti,
Phys.\ Rev.\  D {\bf 69}, 114009 (2004).
%%CITATION = PHRVA,D69,114009;%%

% JLAB data
\bibitem{Malace}
S.~P.~Malace {\it et al.}  [Jefferson Lab E00-116],
Phys.\ Rev.\  C {\bf 80}, 035207 (2009).
%%CITATION = PHRVA,C80,035207;%%

% SLAC data
\bibitem{SLAC}
L.~W.~Whitlow {\it et al.},
Phys.\ Lett.\ B{\bf 282}, 475 (1992).

% NMC data
\bibitem{NMC}
M.~Arneodo {\it et al.},
Nucl.\ Phys.\ B{\bf 483}, 3 (1997).
 
% BCDMS data
\bibitem{BCDMS}
A.~C.~Benvenuti {\it et al.},
Phys.\ Lett.\ B{\bf 223}, 485 (1989); 
{\it ibid.} B{\bf 236}, 592 (1989).

% H1 data
\bibitem{H1}
C.~Adloff {\it et al.},
Eur.\ Phys.\ J.\ C{\bf 19}, 269 (2001);
{\it ibid.} C {\bf 21}, 33 (2001).

% Zeus data
\bibitem{ZEUS}
S.~Chekanov {\it et al.},
Eur.\ Phys.\  J.\ C{\bf 21}, 443 (2001).

% CCFR F2
\bibitem{CCFR2}
U.\ K.\ Yang {\it et al.},
Phys.\ Rev.\ Lett.\ {\bf 86}, 2742 (2001). 

% CCFR xF3
\bibitem{CCFR3}
W.\ G.\ Seligman {\it et al.},
Phys.\ Rev.\ Lett.\ {\bf 79}, 1213 (1997). 

% E605 data
\bibitem{E605}
G.\ Moreno {\it et al.},
Phys.\ Rev.\ D{\bf 43}, 2815 (1991).

% E866 data
\bibitem{E866}
J.~Webb,
Ph.D. Thesis, New Mexico State University (2002),
arXiv:hep-ex/0301031;\\
%
P.~Reimer,
private communication.

%CDF Lasy 98
\bibitem{CDF98}
F.\ Abe {\it et al.},
Phys.\ Rev.\ Lett.\ {\bf 81}, 5754 (1998).

% CDF Lasy 05
\bibitem{CDF05}
D.~ Acosta {\it et al.},
Phys.\ Rev.\ D{\bf71}, 051104(R) (2005).

% D0 Lasy 08
\bibitem{D008}
V.~M.~Abazov {\it et al.},
Phys.\ Rev.\  D{\bf 77}, 011106(R) (2008).

% D0 Lasy e(08)
\bibitem{D0_e08}
V.~M.~Abazov {\it et al.},
Phys.\ Rev.\ Lett.\  {\bf 101}, 211801 (2008).

% CDF W asymmetry
\bibitem{CDF09}
T.~Aaltonen {\it et al.}  [CDF Collaboration],
arXiv:0901.2169 [hep-ex].
%%CITATION = ARXIV:0901.2169;%%

% CDF Jet data (Run I)
\bibitem{CDFjet}
T.~Affolder {\it et al.},
Phys.\ Rev.\ D{\bf 64}, 032001 (2001).

% D0 Jet dta (Run I)
\bibitem{D0jet}
B.~Abbott {\it et al.},
Phys.\ Rev.\ Lett.\ {\bf 86}, 1707 (2001).

% D0 photon + jet
\bibitem{D0gamjet}
V.~M.~Abazov {\it et al.},
Phys.\ Lett.\ B{\bf 666}, 435 (2008).

\bibitem{GL}
F.~Gross and S.~Liuti,
Phys.\ Rev.\  C {\bf 45}, 1374 (1992).
%%CITATION = PHRVA,C45,1374;%%

\bibitem{future}
A.~Accardi {\it et al.}, follow-up paper in preparation.

\bibitem{Arrington}
J.~Arrington, F.~Coester, R.~J.~Holt and T.~S.~Lee,
J.\ Phys.\ G {\bf 36}, 025005 (2009).
%%CITATION = JPHGB,G36,025005;%%

\bibitem{DGG}
A.~De~R\'ujula, H.~Georgi and S.L.~Glashow,
Phys. Rev. D {\bf 12}, 147 (1975).

\bibitem{quarter}
R.~P.~Feynman,
{\em Photon Hadron Interactions}      
(Benjamin, Reading, Massachusetts, 1972);
%
F.~E.~Close,
Phys. Lett. {\bf 43} B, 422 (1973);
%
F.~E.~Close and A.~W.~Thomas,
Phys. Lett. B {\bf 212}, 227 (1988);
%
N.~Isgur,    
Phys. Rev. D {\bf 59}, 034013 (1999). 

\bibitem{FJ}
G.~R.~Farrar and D.~R.~Jackson,
Phys. Rev. Lett. {\bf 35}, 1416 (1975).

\bibitem{Avakian}
H.~Avakian, S.~J.~Brodsky, A.~Deur and F.~Yuan,
Phys.\ Rev.\ Lett.\  {\bf 99}, 082001 (2007).
%%CITATION = PRLTA,99,082001;%%

\bibitem{MINERvA}
D.~Drakoulakos {\it et al.}  [MINERvA Collaboration],
%``Proposal to perform a high-statistics neutrino scattering 
%experiment using a fine-grained detector in the NuMI beam,''
arXiv:hep-ex/0405002; \\
%
L.~Zhu, private communication.

\bibitem{Souder}
P.~A.~Souder,
AIP Conf.\ Proc.\ {\bf 747}, 199 (2005).
%%CITATION = APCPC,747,199;%%

\bibitem{Hobbs}
T.~Hobbs and W.~Melnitchouk,
Phys.\ Rev.\  D {\bf 77}, 114023 (2008).
%%CITATION = PHRVA,D77,114023;%%

\bibitem{A3}
I.~R.~Afnan {\em et al.},
Phys.\ Lett.\ B {\bf 493}, 36 (2000);
%%CITATION = PHLTA,B493,36;%%
%
I.~R.~Afnan {\it et al.},
Phys.\ Rev.\ C {\bf 68}, 035201 (2003).
%%CITATION = PHRVA,C68,035201;%%

\bibitem{BONUS}
Jefferson Lab experiment E03-012,
H.~Fenker, C.~Keppel, S.~Kuhn and W.~Melnitchouk,
spokespersons,
\texttt{http://www.jlab.org/exp\_prog/experiments/summaries/E03-012.ps}.

\end{thebibliography}
\end{document}